\documentclass[twocolumn]{autart}
\usepackage{bbm}
\usepackage{}    

\usepackage{graphicx}          
\usepackage{subfigure}
\usepackage{cite}
\usepackage{amsmath}
\usepackage{amssymb}
\usepackage{mathrsfs}
\usepackage{algorithm}
\usepackage{algorithmicx}
\usepackage{arydshln} 
\usepackage{myhyphen} 
\usepackage{color}
\usepackage{changebar}
\usepackage{booktabs}

\DeclareMathOperator*{\argmin}{\arg\,\min}

\begin{document}

\begin{frontmatter}

\title{Data-Driven Robust Receding Horizon Fault Estimation \thanksref{footnoteinfo}} 

\thanks[footnoteinfo]{This paper was not presented at any IFAC
meeting. Corresponding author Yiming Wan. Tel.: +31152787019;
Fax: +31152786679.}

\author[DCSC]{Yiming Wan}\ead{y.wan@tudelft.nl},    
\author[DCSC]{Tamas Keviczky}\ead{t.keviczky@tudelft.nl},               
\author[DCSC]{Michel Verhaegen}\ead{m.verhaegen@tudelft.nl},  
\author[Linkoping]{Fredrik Gustafsson}\ead{fredrik.gustafsson@liu.se}

\address[DCSC]{Delft Center for Systems and Control, Delft University of Technology, Delft, 2628 CD, The Netherlands}  
\address[Linkoping]{Department of Electrical Engineering, Link\"{o}ping University, SE-581 83 Link\"{o}ping, Sweden}             


\begin{abstract}                          
This paper presents a data-driven receding horizon fault estimation method for additive actuator and sensor faults in unknown linear time-invariant systems, with enhanced robustness to stochastic identification errors. State-of-the-art methods construct fault estimators with identified state-space models or Markov parameters, but they do not compensate for identification errors. Motivated by this limitation, we first propose a receding horizon fault estimator parameterized by predictor Markov parameters. This estimator provides (asymptotically) unbiased fault estimates as long as the subsystem from faults to outputs has no unstable transmission zeros. When the identified Markov parameters are used to construct the above fault estimator, zero-mean stochastic identification errors appear as model uncertainty multiplied with unknown fault signals and online system inputs/outputs (I/O). Based on this fault estimation error analysis, we formulate a mixed-norm problem for the offline robust design that regards online I/O data as unknown. An alternative online mixed-norm problem is also proposed that can further reduce estimation errors when the online I/O data have large amplitudes, at the cost of increased computational burden. Based on a geometrical interpretation of the two proposed mixed-norm problems, systematic methods to tune the user-defined parameters therein are given to achieve desired performance trade-offs. Simulation examples illustrate the benefits of our proposed methods compared to recent literature.
\end{abstract}

\end{frontmatter}

\section{Introduction}
Model-based fault diagnosis techniques for linear dynamic systems have been well established during the past two decades \cite{Blanke2006, ChenPatton1999, Ding2013book, Gust2001}. Recently, the model-based receding horizon approach has received attention because it provides a flexible framework to  enhance robustness of passive fault diagnosis \cite{Wan2013, Zhang2014} and to enable optimal input design in active fault diagnosis \cite{Raim2013a, Raim2013b, Siman2005}. However, an explicit and accurate system model is often unknown in practice. In such situations, a conventional approach first identifies the system model from system I/O data, and then designs the model-based fault diagnosis system under various performance criteria \cite{Simani2003, Patward2005, Manuja2009}. Without explicitly identifying a system model, recent research efforts investigate data-driven approaches to construct a fault diagnosis system utilizing the link between system identification and the model-based fault diagnosis methods \cite{Russell2000, Ding2014book, Ding2014JPC}. These recent data-driven approaches simplify the design procedure by skipping the realization of an explicit system model, while at the same time allow developing systematic methods to address the same fault diagnosis performance criteria as the existing model-based approaches.

Most recent data-driven fault diagnosis approaches for unknown linear dynamic systems can be classified into two categories. The first category, e.g., \cite{QinLi2001} and \cite{Ding2009JPC, Ding2014JPC}, identifies a projection matrix known as parity space/vectors for residual generation, by exploiting the subspace identification method based on principal component analysis (SIM-PCA) \cite{HuangDing2005}.
However, as pointed out in \cite{Dong2012a}, a model reduction step is needed to determine the projection matrix, hence leads to the nonlinear dependence of the generated residuals on the identification errors. Therefore it is difficult to guarantee the robustness of such data-driven methods to the identification errors.

The second category of data-driven fault diagnosis methods, e.g., \cite{Dong2009thesis}, utilizes the Markov parameters (or impulse response parameters) which can be obtained in the first step of the predictor based subspace identification (PBSID) technique \cite{Chiuso2007a, Veen2012}. It constructs residual generators parameterized by the predictor Markov parameters. The main advantage of this method is that the residual signal linearly depends on the identification errors of the predictor Markov parameters. Hence a robust scheme has been developed in \cite{Dong2012a, Dong2012b} to cope with stochastic identification errors. This benefit of robustness compared to the SIM-PCA based method in \cite{Ding2009JPC} is achieved at the cost of increased computational burden in incorporating past I/O data.

Most of the data-driven fault diagnosis literature mentioned above discuss only fault detection and isolation. It is much more involved to estimate/identify the fault signal in the data-driven setting. The work in \cite{Alcala2009, Qin2009} proposed to reconstruct faults by minimizing the reconstructed squared prediction error obtained from PCA. However, this approach did not fully investigate the statistical properties of the calculated fault estimates.
By investigating the link between system-inversion based fault reconstruction and the predictor Markov parameters, the method in \cite{Dong2012c} constructed fault estimators parameterized by the predictor Markov parameters. Its fault estimates are asymptotically unbiased as the estimation horizon length tends to infinity, under the condition that the underlying inverted system is stable.

One drawback of the data-driven fault estimator proposed in \cite{Dong2012c} is that it cannot be directly applied to sensor faults in an unstable open-loop plant because its underlying inverted system is unstable. Another limitation of this method is that it does not compensate for the identification errors. The robustness of fault estimation to the identification errors is critical in two situations: 1) there exist large identification errors due to small number of identification data samples or low signal-to-noise ratio in identification data; 2) multiplication of the erroneous identified matrices with online I/O data of large amplitude cannot be simply ignored.

Motivated by the above two drawbacks of the proposed method in \cite{Dong2012c}, this paper develops data-driven robust fault estimation methods for additive actuator/sensor faults, utilizing the identified Markov parameters. In order to pave the way for data-driven design, we first construct a receding horizon (RH) fault estimator parameterized by the predictor Markov parameters, assuming that the predictor Markov parameters are accurately available. It gives (asymptotically) unbiased fault estimates under the condition that the subsystem from faults to outputs has no unstable transmission zeros. The above condition for unbiasedness generalizes the requirement of stable inversion in \cite{Dong2012c}. An immediate benefit is that our fault estimator can be applied to sensor faults in unstable open-loop plants as long as the above condition for unbiasedness is satisfied, whereas the proposed method in \cite{Dong2012c} cannot.

Our data-driven design parameterizes the above RH fault estimator with predictor Markov parameters identified from closed-loop data. The obtained data-driven fault estimation error is linear with regards to the stochastic identification errors of Markov parameters, although the identification errors appear as multiplicative uncertainty that couples with unknown fault signals as well as online I/O data. In order to enhance robustness to  stochastic identification errors, we propose two mixed-norm fault estimators.
The first one can be designed offline by regarding the online I/O data as unknown. By exploiting online I/O data in its formulated mixed-norm problem, the second robust fault estimator further reduces estimation errors when the online I/O data have large amplitudes, at the cost of increased online computational burden.  Based on a geometric interpretation of the formulated mixed-norm problems, a systematic tuning method for the user-defined parameters therein is provided to achieve the desired trade-offs between estimation bias and variance. 
Our proposed methods can handle sensor and actuator faults either separately or simultaneously. Only the separate scenario is illustrated in detail in this paper. Exact formulas for the simultaneous scenario can be derived in a straightforward manner but are omitted for the sake of brevity.

The rest of this paper starts with the problem formulation and some preliminaries on closed-loop identification of predictor Markov parameters in Section \ref{sect:probformulation}. Section \ref{sect:RHFE_predictor} constructs the predictor-based RH fault estimator, and analyzes its condition for unbiasedness. A data-driven nominal fault estimator is given in Section \ref{sect:RHFE_dd_nominal}. Section \ref{sect:dd_robust} and \ref{sect:dd_robust_onlineopt} propose two mixed-norm fault estimators with enhanced robustness to identification errors. Simulation studies are finally given in Section \ref{sect:sim}.

\section{Preliminaries and problem formulation}\label{sect:probformulation}

\subsection{Notations}
For a matrix $X$, its range and null space is denoted by $\mathcal{R}\left( X \right)$ and $\mathcal{N}\left( X \right)$, respectively. $X^{-}$ represents the left inverse satisfying $X^{-} X = I$, while $X^{(1)}$ represents the generalized inverse satisfying
\begin{equation}\label{eq:ginv}
X X^{(1)} X = X.
\end{equation}
$X^{[i]}$ represents the $i^{\mathrm{th}}$ column of $X$. The trace of $X$ is denoted by $\mathrm{tr}\left( X \right)$. Let $\left\| X \right\|_F$ represent the Frobenius norm of the matrix $X$. The minimal eigenvalue of a symmetric matrix $X$ is represented by $\lambda_{\text{min}} \left( X \right)$. Let $\mathrm{vec}\left( X \right)$ represent the column vector concatenating the columns of a matrix $X$. The symbol ``$\otimes$'' stands for Kronecker product. Let $\mathrm{diag}\left( X_1, X_2, \cdots, X_n \right)$ denote a block diagonal matrix with $X_1, X_2, \cdots, X_n$ as its diagonal matrices.

\subsection{Problem formulation}\label{sect:system_descrip}
We consider linear discrete-time systems governed by the following state space model:
\begin{equation}\label{eq:sys}
\begin{aligned}
\xi (k+1) &= A \xi(k) + Bu(k) + E f(k) + F w(k) \\
y(k) &= C \xi (k) + Du(k) + G f(k) + v(k).
\end{aligned}
\end{equation}
Here $\xi (k)\in\mathbb{R}^{n}$, $y(k)\in\mathbb{R}^{n_y}$, and $u(k)\in\mathbb{R}^{n_u}$ represent the state, the output measurement, and the known control input at time instant $k$, respectively. The process and measurement noises $w(k)\in \mathbb{R}^{n_w}$ and $v(k) \in \mathbb{R}^{n_v}$ are white zero-mean Gaussian, with covariance matrices
$\mathrm{E}\left( w(k) w^\mathrm{T}(k) \right) = Q$,
$\mathrm{E}\left( v(k) v^\mathrm{T}(k) \right) = R$,
$\mathrm{E}\left( w(k) v^\mathrm{T}(k) \right) = 0$.
$f(k)\in\mathbb{R}^{n_f}$ is the unknown fault signal to be estimated.
$A, B, C, D, E, F, G$ are constant real matrices, with bounded norms and appropriate dimensions.

The following assumption is standard in Kalman filtering \cite{Kailath2000} and subspace identification \cite{Chiuso2007a, Kata2005}:
\begin{assum}\label{ass:detect_control}
The pair $\left( C, A \right)$ is assumed detectable; and there are no uncontrollable modes of $\left( A, F Q^{\frac{1}{2}} \right)$ on the unit circle, where $Q^{\frac{1}{2}} \cdot \left( Q^{\frac{1}{2}} \right)^\mathrm{T} = Q$ is the covariance matrix of $w(k)$.
\end{assum}

Based on Assumption \ref{ass:detect_control}, the system (\ref{eq:sys}) admits the one-step-ahead predictor form given by \cite{Kailath2000}
\begin{equation}\label{eq:predictor}
\begin{aligned}
x(k+1) &= \Phi x(k) + \tilde B u(k) + \tilde E f(k) + K y(k) \\
y(k) &= C x(k) + D u(k) + G f(k) + e(k),
\end{aligned}
\end{equation}
where $K$ is the steady-state Kalman gain, $\Phi = A - KC$, $\tilde B = B - KD$, and $\tilde E = E - KG$, $\left\{e(k)\right\}$ is the zero-mean innovation process with the covariance matrix $\Sigma_e$.

We consider additive sensor or actuator faults in this paper, i.e.,
\begin{itemize}
\item fault of the $j^{th}$ sensor: 
	\begin{equation}\label{eq:senfault_model}
	E = 0_{n_x \times 1}, \; G = I^{[j]},\; \tilde E = - K^{[j]};
	\end{equation}
\item fault of the $l^{th}$ actuator: 
	\begin{equation}\label{eq:actfault_model}
	E = B^{[l]}, \; G = D^{[l]},\; \tilde E = \tilde B^{[l]};
	\end{equation}
\item simultaneous faults of the $j^{th}$ sensor and $l^{th}$ actuator: 
	\begin{equation}\label{eq:senactfault_model}
	E = \left[\begin{matrix}
	0_{n_x \times 1} & B^{[l]}
	\end{matrix}\right], 
	G = \left[ \begin{matrix}
	I^{[j]} & D^{[l]}
	\end{matrix} \right], 
	\tilde E = \left[\begin{matrix}
	- K^{[j]} & \tilde B^{[l]}
	\end{matrix}\right];
	\end{equation}
\end{itemize}
with $X^{[j]}$ representing the $j^{\mathrm{th}}$ column of a matrix $X$. 

Denote the predictor Markov parameters by
\begin{equation}\label{eq:markov_param}
\begin{aligned}
& H_i^u = \left\{ \begin{array}{ll}
D & i=0 \\
C \Phi^{i-1} \tilde B & i>0
\end{array} \right. , \;
H_i^y = \left\{ \begin{array}{ll}
0 & i=0 \\
C \Phi^{i-1} K & i>0
\end{array} \right. , \\
& H_i^f = \left\{ \begin{array}{ll}
G & i=0 \\
C \Phi^{i-1} \tilde E & i>0
\end{array} \right. .
\end{aligned}
\end{equation}
\vspace*{-0.5cm}

\begin{assum}\label{ass:fault_rank}
	The relative degree of the fault subsystem $\left(\Phi, \tilde E, C, G\right)$ is $\tau$, i.e., $\tau$ is the smallest nonnegative integer $i$ such that $H_0^f = H_1^f = \cdots = H_{i-1}^f = 0$ and $H_i^f \neq 0$ \cite{Kirt2011}; moreover, 
	$\mathrm{rank}\left(H_\tau^f\right) = n_f$ \cite{Dong2012c}.
\end{assum}
Note that $\tau=0$ for sensor faults and $\tau > 0$ for actuator faults.

%


The essential goals of this paper are to design a fault estimator from identification data without knowing the system matrices in (\ref{eq:sys}), and moreover to robustify the fault estimator against identification errors.

Concerning the identification data, it should be noted that in practice data from faulty conditions may be seldomly available, or if recorded then without a reliable fault description \cite{Ding2014JPC}. Hence we make the assumption as below:
\begin{assum}\label{ass:data}
Only I/O data collected from the fault-free condition are used in our data-driven design.
\end{assum}

In contrast to \cite{Park2000} which assumes the fault signals $f(k)$ evolve according to a random walk model, no assumption is made in this paper about how the fault signals $f(k)$ vary with time.

\subsection{Closed-loop identification of predictor Markov parameters}\label{sect:IDmarkov}
Considering Assumption \ref{ass:data}, we set $f(k) = 0$ in (\ref{eq:sys}) for the identification data collected from the fault-free condition.
Then with $f(k) = 0$, the predictor form (\ref{eq:predictor}) over the time window $\left[ t, \cdots, t+N-1 \right]$ can be written into the following data equation \cite{Chiuso2007a, Veen2012}:
\begin{equation}\label{eq:data_eq}
\mathbf{Y}_{\mathrm{id}} = C \Phi^p \mathbf{X}_{\mathrm{id}} + \Xi \mathbf{Z}_{\mathrm{id}} + \mathbf{E}_{\mathrm{id}},
\end{equation}
where
\begin{align}
\Xi &= \left[ \begin{array}{cccccc}
				H_{p}^u & H_{p}^y & \cdots & H_{1}^u & H_{1}^y & H_{0}^u
				\end{array}
			\right]   \label{eq:Xi}
\end{align}
denotes the sequence of Markov parameters $\{H_i^u\}$ and $\{H_i^y\}$ (defined in (\ref{eq:markov_param})) to be identified. 
The detailed definitions of the data matrices $\mathbf{X}_{\rm{id}}$, $\mathbf{Y}_{\mathrm{id}}$ and $\mathbf{Z}_{\mathrm{id}}$ can be found in \cite{Veen2012}, and $\mathbf{E}_{\rm{id}}$ is the sequence of the innovation signal in the identification data.

The least-squares (LS) estimate of the Markov parameters $\Xi$ is
\begin{equation}\label{eq:LS_id}
\begin{aligned}
\hat \Xi &= \argmin\limits_{\Xi}
\left\| \mathbf{Y}_{\mathrm{id}} - \Xi \mathbf{Z}_{\mathrm{id}}  \right\|_F^2
= \mathbf{Y}_{\mathrm{id}} \mathbf{Z}_{\mathrm{id}}^{-} \\
&= \Xi + C \Phi^p \mathbf{X}_{\mathrm{id}} \mathbf{Z}_{\mathrm{id}}^{-}
+ \mathbf{E}_{\mathrm{id}} \mathbf{Z}_{\mathrm{id}}^{-},
\end{aligned}
\end{equation}
with $\mathbf{Z}_{\mathrm{id}}^{-} =
\mathbf{Z}_{\mathrm{id}}^{\mathrm{T}}
\left( \mathbf{Z}_{\mathrm{id}} \mathbf{Z}_{\mathrm{id}}^{\mathrm{T}} \right)^{-1}$.
As standard assumptions for consistent identification from closed-loop data, we assume that 1) the data matrix $\mathbf{Z}_{\mathrm{id}}$ has full row rank, and 2) either the controller has at least one-step delay or the plant model has no direct feedthrough ($D=0$) \cite{Chiuso2007a, Veen2012}.

With sufficiently large $p$, the estimation bias
$C \Phi^p \mathbf{X}_{\mathrm{id}} \mathbf{Z}_{\mathrm{id}}^{-}$ can be neglected. Then the stochastic identification errors are
\begin{equation}\label{eq:Xi_err}
\Delta {\hat \Xi} = \hat \Xi - \Xi \approx \mathbf{E}_{\mathrm{id}} \mathbf{Z}_{\mathrm{id}}^{-}.
\end{equation}
Hence according to (\ref{eq:Xi_err}), the identification errors in Markov parameters can also be written as
\begin{equation}\label{eq:iderr_markov}
\begin{aligned}
\Delta H_i^u &= \hat H_i^u - H_i^u = \mathbf{E}_{\mathrm{id}} {M}_i^u, \\
\Delta H_i^y &= \hat H_i^y - H_i^y = \mathbf{E}_{\mathrm{id}} {M}_i^y,
\end{aligned}
\end{equation}
where $\hat H_i^u$ and $\hat H_i^y$ represent the estimated Markov parameters in $\hat \Xi$ given by (\ref{eq:LS_id}), ${M}_i^u$ and ${M}_i^y$ are the corresponding blocks of $\mathbf{Z}_{\mathrm{id}}^{-}$, i.e.,
\begin{equation}\label{eq:Muy}
\mathbf{Z}_{\mathrm{id}}^{-} =
\left[ \begin{array}{cccccc}
               M_{p}^u & M_{p}^y & \cdots & M_{1}^u & M_{1}^y & M_{0}^u
             \end{array}
 \right],\;
M_{0}^y = 0.
\end{equation}

The innovation covariance can be estimated by \cite{Gust2001, Kata2005}
\begin{equation}\label{eq:hat_Sigma_e}
\hat \Sigma_e = \mathrm{cov}
\left( \mathbf{Y}_{\mathrm{id}} - \hat \Xi \mathbf{Z}_{\mathrm{id}} \right).
\end{equation}
For the sake of brevity, we shall not distinguish between the estimated innovation covariance $\hat \Sigma_e$ and its true value $\Sigma_e$ in the rest of this paper.

\section{Predictor-based receding horizon fault estimation}\label{sect:RHFE_predictor}
In this section, we will construct an RH fault estimator based on the predictor form of the system (\ref{eq:sys}). Here we consider the predictor form instead of the original system model (\ref{eq:sys}) in order to pave the way for data-driven design. 


Consider a sliding window with a length of $L$ sampling instants. Define stacked data vectors in this window as $\mathbf{u}_{k,L}$, $\mathbf{y}_{k,L}$, $\mathbf{f}_{k,L}$, and $\mathbf{e}_{k,L}$, respectively for the signals $u$, $y$, $f$, and $e$; e.g.,
\begin{equation}\label{eq:ukL}
\mathbf{u}_{k,L} = \left[ \begin{array}{ccc}
                                  u^\mathrm{T}\left(k_0\right) & \cdots & u^\mathrm{T}\left(k\right)
                                \end{array} \right]^\mathrm{T},
\end{equation}
with $k_0 = k-L+1$. For the predictor form  (\ref{eq:predictor}), let $\mathcal{O}_L$ denote its extended observability matrix with $L$ block elements, and $\mathbf{T}_{L}^{\star}$ be the lower triangular block-Toeplitz matrix with $L$ block columns and rows, with $\star$ representing $u$, $y$, or $f$:
\begin{equation}\label{eq:OL_TLu}
\mathcal{O}_{L} = \left[ \begin{matrix}
                          C \\
                          C \Phi \\
                          \vdots \\
                          C \Phi^{L-1}
                        \end{matrix} \right], \;
\mathbf{T}_{L}^{\star} = \left[ \begin{matrix}
                          H_0^\star & 0 & \ldots & 0 \\
                          H_1^\star & H_0^\star & \ddots & \vdots \\
                          \vdots & \vdots & \ddots & 0  \\
                          H_{L-1}^\star & H_{L-2}^\star & \cdots & H_0^\star
                        \end{matrix} \right].
\end{equation}

Given the I/O data over the sliding window $\left[ k_0, k \right]$,
the stacked residual signal $\mathbf{r}_{k,L}$ in $\left[ k_0, k \right]$ can be computed by
\begin{equation}\label{eq:resL_compute}
\mathbf{r}_{k,L} = \mathbf{y}_{k,L} - \mathbf{T}_L^y \mathbf{y}_{k,L} - \mathbf{T}_L^u \mathbf{u}_{k,L},
\end{equation}
according to the predictor form (\ref{eq:predictor}).
We can further write down the transitions from unknown initial state, faults and noises to the stacked residual signal $\mathbf{r}_{k,L}$ as
\begin{equation}\label{eq:res_dyn_batch0}
\mathbf{r}_{k,L} = \mathcal{O}_{L} x(k_0) + \mathbf{T}_{L}^{f} \mathbf{f}_{k,L} + \mathbf{e}_{k,L}.
\end{equation}
With Assumption \ref{ass:fault_rank}, (\ref{eq:res_dyn_batch0}) can be simplified as
\begin{equation}\label{eq:res_dyn_batch}
\begin{aligned}
\mathbf{r}_{k,L}
&= \underbrace{\left[ \begin{array}{cc}
                        \mathcal{O}_{L} & \mathbf{T}_{L,\tau}^{f}
                      \end{array} \right]}_{\Psi_{L,\tau}} 
   \underbrace{\left[ \begin{array}{c}
                        x(k_0) \\
                        \mathbf{f}_{k-\tau,L-\tau}
                      \end{array} \right]}_{\mathbf{f}_{k-\tau,L-\tau}^x} + \mathbf{e}_{k,L},
\end{aligned}
\end{equation}
where $\tau$ is the relative degree of the fault subsystem $\left( A, E, C, G \right)$, $\mathbf{T}_{L,\tau}^{f}$ represents the first $L-\tau$ block-columns of $\mathbf{T}_{L}^{f}$ defined similar to (\ref{eq:OL_TLu}), $\mathbf{f}_{k-\tau,L-\tau}$ is defined in the same way as in (\ref{eq:ukL}).

With (\ref{eq:res_dyn_batch}), we can formulate the receding horizon fault estimation (RHFE) problem
\begin{equation}\label{eq:LS_prob}
\mathop {\min }\limits_{{{\mathbf{{{f}}}}_{k-\tau,L-\tau}^x}} \;\left\| {{{\bf{r}}_{k,L}} - \Psi_{L,\tau}
\mathbf{{{f}}}_{k-\tau,L-\tau}^x } \right\|_{\Sigma_{e,L}^{-1}}^2
\end{equation}
in the LS sense, with
\begin{equation}\label{eq:Sigma_eL}
\Sigma_{e,L} = I_L \otimes \Sigma_e
\end{equation}
denoting the covariance matrix of $\mathbf{e}_{k,L}$.
It has non-unique solutions because $\Psi_{L,\tau}$ may not have full column rank. One solution to the problem (\ref{eq:LS_prob}) is
\begin{equation}\label{eq:fkL_hat}
{{\mathbf{\hat {{f}}}}_{k-\tau,L-\tau}^x} = \left( \Psi_{L,\tau}^\mathrm{T} \Sigma_{e,L}^{-1} \Psi_{L,\tau} \right)^{(1)}
\Psi_{L,\tau}^\mathrm{T} \Sigma_{e,L}^{-1} {{\bf{r}}_{k,L}}.
\end{equation}
We will show in the following theorem, however, that the last $n_f$ entries of ${{\mathbf{\hat {{f}}}}_{k-\tau,L-\tau}^x}$, i.e.,
\begin{equation}\label{eq:f_hat_tao}
\begin{aligned}
\hat f \left( k - \tau \right) = \mathcal{I}_{n_f} {{\mathbf{\hat {{f}}}}_{k-\tau,L-\tau}^x}
\end{aligned}
\end{equation}
with $\mathcal{I}_{n_f} = [ \begin{array}{cc}
                        0 & I_{n_f}
                      \end{array}] \in \mathbb{R}^{n_f \times \left( n + n_f \left( L - \tau \right) \right)}$, 
represent an (asymptotically) unbiased estimate of $f \left( k - \tau \right)$ under certain conditions. The estimation delay $\tau$ in (\ref{eq:f_hat_tao}) is caused by the relative degree in Assumption \ref{ass:fault_rank}.

\begin{thm}\label{thm:unbias}
Let $\tau$ and $\nu$ denote the relative degree and the observability index of the fault subsystem $( \Phi, \tilde E, C, G )$, respectively.
\begin{enumerate}
  \item[(\romannumeral1)] The $\tau$-delay fault estimate $\hat f(k-\tau)$ defined in (\ref{eq:f_hat_tao}) is unbiased for all $L \geq \nu+\tau$ if and only if $( \Phi, \tilde E, \mathcal{O}_{\tau+1}, \mathbf{H}_{\tau}^f )$ has no transmission zeros, with 
\begin{equation}\label{eq:Htauf}
\mathbf{H}_{\tau}^f =
\left[ \begin{array}{cccc}
          ( H_{0}^f )^\mathrm{T} & ( H_{1}^f )^\mathrm{T} & \cdots & ( H_{\tau}^f )^\mathrm{T}
       \end{array}
 \right]^\mathrm{T}.
\end{equation}
  \item[(\romannumeral2)] The $\tau$-delay fault estimate $\hat f(k-\tau)$ is asymptotically unbiased for $L \rightarrow \infty$ if and only if all transmission zeros of $( \Phi, \tilde E, \mathcal{O}_{\tau+1}, \mathbf{H}_{\tau}^f )$ are stable.
\end{enumerate}
\end{thm}

The proof is given in Appendix \ref{app:thm_unbias}.

Instead of including the unknown initial state as in the RHFE problem (\ref{eq:LS_prob}), the essential idea of \cite{Dong2012c} is to find a lower triangular block-Toeplitz matrix $\mathbf{T}_L^g$ such that $\mathbf{T}_L^g \cdot \mathbf{T}_{L,\tau}^f = I$ and the estimation error caused by the unknown initial state exponentially decays with $L$. The condition for unbiasedness in \cite{Dong2012c} requires that the inverse system related to $\mathbf{T}_L^g$ is stable. However this has several drawbacks: it does not clarify how the unbiasedness condition is related to the system property of the underlying plant; and moreover, for the case of sensor faults in an open-loop unstable plant, the method in \cite{Dong2012c} cannot find a stable left inverse matrix $\mathbf{T}_L^g$ for $\mathbf{T}_{L,\tau}^f$.

On the contrary, Theorem \ref{thm:unbias} clearly states that the condition for unbiasedness is related to the invariant zeros of the fault subsystem in the underlying plant. An immediate benefit is that our proposed RH fault estimator can ensure (asymptotically) unbiased estimates for sensor faults in an open-loop unstable plant, as long as the fault subsystem has no unstable transmission zeros.

\begin{rem}
The unbiasedness condition of the $\tau$-delay fault estimate stated in Theorem \ref{thm:unbias} has close links with the $\tau$-delay left inversion in \cite{Massey1968, Gill2007} and the $\tau$-delay input and initial-state reconstruction in \cite{Kirt2011}. However, the $\tau$-delay left inversion in \cite{Massey1968, Gill2007} requires the initial state to be known a priori, while the $\tau$-delay input and initial-state reconstruction in \cite{Kirt2011} requires observability of the pair $\left( \Phi, C \right)$ to simultaneously reconstruct the initial state with the unknown input. Although it seems that the RHFE problem (\ref{eq:LS_prob}) jointly estimates initial state and faults, we are actually only interested in the fault estimate without unbiased reconstruction of the unknown initial state. This is an intuitive reason why Theorem \ref{thm:unbias} can cope with the unknown initial state in the case that $\left( \Phi, C \right)$ is detectable.
\end{rem}

\begin{rem}
Theorem \ref{thm:unbias} above generalizes Theorems 1 and 2 in \cite{Wan2014} in two aspects: 1) Theorems 1 and 2 in \cite{Wan2014} are limited to the case $\tau=0$, while Theorem \ref{thm:unbias} here applies to general relative degrees; 2) Theorems 1 and 2 in \cite{Wan2014} focus on the fault estimator constructed with the original system (\ref{eq:sys}), while in this work we construct in Theorem \ref{thm:unbias} the fault estimator with the predictor (\ref{eq:predictor}).
\end{rem}

It should be noted that an RHFE problem similar to (\ref{eq:LS_prob}) can also be formulated using the original system (\ref{eq:sys}), see \cite{Wan2014}. Its equivalence to our RHFE problem (\ref{eq:LS_prob}) is shown in the following theorem.

\begin{thm}\label{thm:solution_equivalent}
If both the original system model (\ref{eq:sys}) and its predictor form (\ref{eq:predictor}) are accurately available, the $\tau$-delay fault estimate $\hat f(k-\tau)$, computed by (\ref{eq:fkL_hat}) and (\ref{eq:f_hat_tao}) based on the predictor form (\ref{eq:predictor}), is equivalent to the fault estimate proposed as Equation (15) in \cite{Wan2014} based on the original system model (\ref{eq:sys}).
\end{thm}

The proof of Theorem \ref{thm:solution_equivalent} is given in Appendix \ref{app:equivalence}. 
The above equivalence implies that the predictor gain $K$ does not affect the statistics of the fault estimation error, and the condition of unbiasedness in Theorem \ref{thm:unbias} holds for the RH fault estimation using the original form.

\section{Data-driven nominal receding horizon fault estimator}\label{sect:RHFE_dd_nominal}
In this section, we will parameterize the RH fault estimator introduced in Section \ref{sect:RHFE_predictor} with the predictor Markov parameters, and then provide the nominal data-driven design method without considering identification errors.

In order to construct the LS fault estimator (\ref{eq:fkL_hat}), we first need to construct the block-Toeplitz matrices $\mathbf{T}_L^u$, $\mathbf{T}_L^y$, and $\mathbf{T}_L^f$ from the predictor Markov parameters according to (\ref{eq:OL_TLu}).
Then, we need the extended observability matrix $\mathcal{O}_L$. One possible approach is to identify $\mathcal{O}_L$ from the block-Hankel matrix
\begin{equation}\label{eq:HLm}
\mathbf{H}_{L,m}^o = \left[
\begin{matrix}
  H_1^u & H_2^u & \cdots & H_m^u \\
  H_2^u & H_3^u & \cdots & H_{m+1}^u \\
  \vdots & \vdots & \ddots & \vdots \\
  H_L^u & H_{L+1}^u & \cdots & H_{L+m-1}^u
\end{matrix}
 \right]
\end{equation}
through a model reduction step \cite{Veen2012}. But this model reduction step would make the fault estimation error depend nonlinearly on 
the identification errors. In order to avoid this difficulty, we substitute 
$\mathcal{O}_L x(k_0) = \mathbf{H}_{L,m}^o \zeta_{m}$ into (\ref{eq:res_dyn_batch}) by exploiting the following property:
\begin{equation}\label{eq:range_HLm}
\mathcal{R} \left( \mathcal{O}_L \right) = \mathcal{R} \left( \mathbf{H}_{L,m}^o \right)
\end{equation} for $m \geq n$.
Then (\ref{eq:res_dyn_batch}) can be rewritten as
\begin{equation}\label{eq:res_batch_markov}
\mathbf{r}_{k,L} = \underbrace{\left[ \begin{array}{cc}
                        \mathbf{H}_{L,m}^o & \mathbf{T}_{L,\tau}^{f}
                      \end{array} \right]}_{\Upsilon_{L,\tau}} 
   \underbrace{\left[ \begin{array}{c}
                        \zeta_m \\
                        \mathbf{f}_{k-\tau,L-\tau}
                      \end{array} \right]}_{\mathbf{f}_{k-\tau,L-\tau}^\zeta} + {\mathbf{e}_{k,L}},
\end{equation}
where $\mathbf{T}_{L,\tau}^{f}$ consists of the first $L-\tau$ block-columns of $\mathbf{T}_{L}^{f}$ defined in (\ref{eq:OL_TLu}).
By doing so, the fault estimation error becomes linear with regards to the identification errors, as shown later in (\ref{eq:fest_err}).
Based on (\ref{eq:res_batch_markov}), an LS problem similar to (\ref{eq:LS_prob}) can be formulated, and one solution is
\begin{equation}\label{eq:hat_fL_markov}
{{\mathbf{\hat {{f}}}}_{k-\tau,L-\tau}^\zeta} =
\left( \Upsilon_{L,\tau}^\mathrm{T} \Sigma_{e,L}^{-1} \Upsilon_{L,\tau} \right)^{(1)} \Upsilon_{L,\tau}^\mathrm{T} \Sigma_{e,L}^{-1} {{\bf{r}}_{k,L}}.
\end{equation}
Similarly to (\ref{eq:f_hat_tao}), we obtain the fault estimate
\begin{equation}\label{eq:f_hat_tao_aa}
\begin{aligned}
\hat f \left( k - \tau \right) = \mathcal{I}_{n_f} {{\mathbf{\hat {{f}}}}_{k-\tau,L-\tau}^\zeta},
\end{aligned}
\end{equation}
with $\mathcal{I}_{n_f} = \left[ \begin{array}{cc}
                                   0 & I_{n_f}
                                 \end{array}
 \right] \in \mathbb{R}^{n_f \times \left( n_u \cdot m + n_y \left( L - \tau \right) \right)}$.

\begin{thm}\label{thm:unbias_markov}
The sufficient and necessary condition for unbiased estimation in Theorem \ref{thm:unbias} applies to the fault estimate defined in (\ref{eq:hat_fL_markov})-(\ref{eq:f_hat_tao_aa}).
\end{thm}

The proof is given in Appendix \ref{app:unbias_markov}.

Combining (\ref{eq:resL_compute}), (\ref{eq:hat_fL_markov}), and (\ref{eq:f_hat_tao_aa}) yields the RH fault estimator as below:
\begin{align}
\hat f(k-\tau) &= \mathcal{G}_\mathrm{n} \mathbf{r}_{k,L} = \mathcal{G}_\mathrm{n} \left[ \begin{array}{c:c}
         I - \mathbf{T}_L^y & -\mathbf{T}_L^u
       \end{array}
 \right] 
 \left[ \begin{array}{c}
          \mathbf{y}_{k,L} \\
          \mathbf{u}_{k,L}
        \end{array}
  \right], \label{eq:RHFEor} \\
\mathcal{G}_\mathrm{n} &=
{\mathcal{I}_{n_f} \left( \Upsilon_{L,\tau}^\mathrm{T} \Sigma_{e,L}^{-1} \Upsilon_{L,\tau} \right)^{(1)} \Upsilon_{L,\tau}^\mathrm{T} \Sigma_{e,L}^{-1} }, \label{eq:RHFEor_F}
\end{align}
where $\mathcal{G}_\mathrm{n}$ represents the nominal RH fault estimator based on the residual signal $\mathbf{r}_{k,L}$.

Without considering the identification errors, the data-driven design of nominal RH fault estimator can now be summarized in Algorithm \ref{alg:nominal_design}. For the sake of brevity, we do not list the estimated fault Markov parameters $\hat H_i^f$ and their estimation errors for simultaneous sensor and actuator faults, because they can be straightforwardly derived similarly to (\ref{eq:hat_Hif_sen}) and (\ref{eq:hat_Hif_act}). Thus all our proposed algorithms in this paper can be directly extended to deal with simultaneous sensor and actuator faults.

\begin{algorithm}
  \caption{Data-driven nominal RH fault estimation}
  \label{alg:nominal_design}
  \begin{algorithmic}
    \State
    \begin{enumerate}
      \item[1)] Collect identification data from the fault-free condition, and form the data matrices $\mathbf{Y}_{\mathrm{id}}$ and $\mathbf{Z}_{\mathrm{id}}$ with sufficiently large $p$ \cite{Veen2012}.
      \item[2)] Compute the sequence of Markov parameters $\hat \Xi$ and the innovation covariance $\hat \Sigma_e$ via  (\ref{eq:LS_id}) and (\ref{eq:hat_Sigma_e}); extract the identified Markov parameters $\hat H_i^u$ and $\hat H_i^y$ from $\hat \Xi$ according to (\ref{eq:Xi}); and extract $\hat H_i^f$ according to (\ref{eq:senfault_model})-(\ref{eq:markov_param}):
			\begin{itemize}
				\item for $j^{\text{th}}$ sensor faults:
					\begin{equation}\label{eq:hat_Hif_sen}
						\hat H_i^f = - ( \hat H_i^{y} )^{[j]} \;\mathrm{for}\; i > 0, \;\mathrm{and}\; \hat H_0^f = I^{[j]};
					\end{equation}
				\item or for $l^{\text{th}}$ actuator faults:
					\begin{equation}\label{eq:hat_Hif_act}
						\hat H_i^f = ( \hat H_i^{u} )^{[l]} \;\; (i \geq 0).
					\end{equation}
			\end{itemize}
      \item[3)] Select sufficiently large $L$.
          Construct the estimates of $\Sigma_{e,L}$ in (\ref{eq:Sigma_eL}), $\mathbf{T}_L^y$, $\mathbf{T}_L^u$, $\mathbf{T}_L^f$ in (\ref{eq:OL_TLu}), $\mathbf{H}_{L,m}^o$ in (\ref{eq:HLm}), and ${\Upsilon}_{L,\tau}$ in (\ref{eq:res_batch_markov}) as $\hat \Sigma_{e,L}$, $\mathbf{\hat T}_L^y$, $\mathbf{\hat T}_L^u$, $\mathbf{\hat T}_{L}^f$, $\mathbf{\hat H}_{L,m}^o$, and $\hat {\Upsilon}_{L,\tau}$ by using
          $\hat \Sigma_e$ and the identified Markov parameters $\{ \hat H_i^u, \hat H_i^y, \hat H_i^f \}$. Form $\mathbf{\hat T}_{L,\tau}^f$ with the first $L-\tau$ block-columns of $\mathbf{\hat T}_{L}^f$.
      \item[4)] Compute the nominal fault estimator according to (\ref{eq:RHFEor}) and (\ref{eq:RHFEor_F}).
    \end{enumerate}
  \end{algorithmic}
\end{algorithm}

\section{Data-driven robust receding horizon fault estimation}\label{sect:dd_robust}
The data-driven nominal design in Algorithm \ref{alg:nominal_design} might give biased fault estimates due to errors in the identified Markov parameters. To address this problem, this section proposes an offline robust design which regards the online I/O data as unknown in the design stage.

\subsection{Data-driven robust design}
Since the Markov parameters related to faults are extracted from $\hat H_i^u$ or $\hat H_i^y$ via (\ref{eq:hat_Hif_act}) or (\ref{eq:hat_Hif_sen}), the identification errors of $\hat H_i^f$ can be  expressed as
\begin{equation}\label{eq:delta_Hif}
\Delta H_i^f = \mathbf{E}_{\mathrm{id}} {M}_i^f,
\end{equation}
where
\begin{equation}\label{eq:Mif}
{M}_i^f = \left\{ \begin{array}{ll}
                    \left( {M}_i^u \right)^{[j]} & \mathrm{for\; faults\; of\; the} \; j^{\mathrm{th}} \;\mathrm{actuator} \\
                    -\left( {M}_i^y \right)^{[j]} & \mathrm{for\; faults\; of\; the} \; j^{\mathrm{th}} \;\mathrm{sensor}
                  \end{array}
 \right.
\end{equation}
with ${M}_i^u$ and ${M}_i^y$ defined in (\ref{eq:iderr_markov})-(\ref{eq:Muy}).

With (\ref{eq:iderr_markov}) and (\ref{eq:delta_Hif}), the estimated matrices $\mathbf{\hat T}_L^y$, $\mathbf{\hat T}_L^u$, $\mathbf{\hat T}_{L,\tau}^f$, $\mathbf{\hat H}_{L,m}^o$ and $\hat {\Upsilon}_{L,\tau}$ in Algorithm \ref{alg:nominal_design} can be written as
\begin{gather}
\mathbf{\hat H}_{L,m}^o = \mathbf{H}_{L,m}^o + \mathbf{\bar E}_{\mathrm{id}}
\mathbf{\bar M}_{L,m}^o, \;
\mathbf{\hat T}_L^y = \mathbf{T}_L^y - \mathbf{\bar E}_{\mathrm{id}}
 \mathbf{\bar M}_L^y, \label{eq:hatHLm} \\
\mathbf{\hat T}_L^u = \mathbf{T}_L^u + \mathbf{\bar E}_{\mathrm{id}}
 \mathbf{\bar M}_L^u, \;
\mathbf{\hat T}_{L,\tau}^f = \mathbf{T}_{L,\tau}^f + \mathbf{\bar E}_{\mathrm{id}}
 \mathbf{\bar M}_{L,\tau}^f, \\
\hat {\Upsilon}_{L,\tau} = {\Upsilon}_{L,\tau} + \mathbf{\bar E}_{\mathrm{id}} \mathbf{\bar M}_{\Upsilon},
\label{eq:hatPsiLtau}
\end{gather}
where $\mathbf{\bar M}_{L,m}^o$ is the block-Hankel matrix constructed with $M_1^u, M_2^u, \cdots, M_{L+m-1}^u$ similarly to $\mathbf{H}_{L,m}^o$ in (\ref{eq:HLm}), $\mathbf{\bar M}_{L}^\star$ is the block-Toeplitz matrix constructed with $M_0^\star, M_1^\star, \cdots, M_{L-1}^\star$ similarly to $\mathbf{T}_{L}^\star$ in (\ref{eq:OL_TLu}) with $\star$ representing $u$, $y$, or $f$,
\begin{equation}\label{eq:Eidbar}
\mathbf{\bar E}_{\mathrm{id}} = \mathrm{diag}
 \underbrace {\left( \mathbf{E}_{\mathrm{id}}, \mathbf{E}_{\mathrm{id}}, \cdots, \mathbf{E}_{\mathrm{id}} \right)}_{L\;blocks},
\end{equation}
\begin{equation}\label{eq:Mbarf}
\mathbf{\bar M}_{\Upsilon} = \left[ \begin{array}{cc}
                                 \mathbf{\bar M}_{L,m}^o  & \mathbf{\bar M}_{L,\tau}^f
                                \end{array} \right],
\end{equation}
and $\mathbf{\bar M}_{L,\tau}^f$ consists of the first $L-\tau$ block-columns of $\mathbf{\bar M}_{L}^f$.

Based on (\ref{eq:hatHLm})-(\ref{eq:hatPsiLtau}), we can write down the residual signal $\mathbf{\hat r}_{k,L}$ considering identification errors according to (\ref{eq:resL_compute})-(\ref{eq:res_dyn_batch}) and (\ref{eq:res_batch_markov}):
\begin{equation}\label{eq:hatrkL}
\begin{aligned}
\mathbf{\hat r}_{k,L} =&\, \mathbf{y}_{k,L} - \mathbf{\hat T}_L^y \mathbf{y}_{k,L} - \mathbf{\hat T}_L^u \mathbf{u}_{k,L} \\
=&\, {\Upsilon}_{L,\tau} \mathbf{f}_{k-\tau,L-\tau}^\zeta + \mathbf{e}_{k,L} + \left( \mathbf{T}_L^y - \mathbf{\hat T}_L^y \right)\mathbf{y}_{k,L} \\
 &\, + \left( \mathbf{T}_L^u - \mathbf{\hat T}_L^u \right)\mathbf{u}_{k,L} \\
=&\, \left( \hat {\Upsilon}_{L,\tau} - \mathbf{\bar E}_{\mathrm{id}} \mathbf{\bar M}_{\Upsilon} \right) \mathbf{f}_{k-\tau,L-\tau}^\zeta
+ \mathbf{e}_{k,L} \\
&\, - \mathbf{\bar E}_{\mathrm{id}}
\underbrace {\left[ \begin{array}{cc}
                      -\mathbf{\bar M}_L^y & \mathbf{\bar M}_L^u
                    \end{array}
 \right]}_{\mathbf{\bar M}_L^z}
\underbrace {\left[ \begin{array}{c}
                      \mathbf{y}_{k,L} \\
                      \mathbf{u}_{k,L}
                    \end{array}
 \right]}_{\mathbf{z}_{k,L}}.
\end{aligned}
\end{equation}

Similarly to $\mathcal{G}_\text{n}$ in (\ref{eq:RHFEor}), let the matrix $\mathcal{G}$ denote the $\tau$-delay fault estimator based on the residual $\mathbf{\hat r}_{k,L}$, i.e.,
\begin{equation}\label{eq:G_rkL}
\hat f(k-\tau) = \mathcal{G} \mathbf{\hat r}_{k,L}.
\end{equation}
It follows from (\ref{eq:hatrkL}) that the fault estimation error is 
\begin{equation}\label{eq:fest_err}
\begin{aligned}
\Delta f(k-\tau) =&\, \hat f(k-\tau) - \mathcal{I}_{n_f} \mathbf{f}_{k-\tau,L-\tau}^\zeta \\
=&\, \underbrace {\left( \mathcal{G} \hat {\Upsilon}_{L,\tau} - \mathcal{G} \mathbf{\bar E}_{\mathrm{id}} \mathbf{\bar M}_{\Upsilon} - \mathcal{I}_{n_f} \right)}_{\mathcal{T}_{f} \left( \mathcal{G} \right)} \mathbf{f}_{k-\tau,L-\tau}^\zeta \\
&\, - \underbrace {\mathcal{G} \mathbf{\bar E}_{\mathrm{id}} {\mathbf{\bar M}_L^z}}_{\mathcal{T}_{z} \left( \mathcal{G} \right)} \mathbf{z}_{k,L} + \mathcal{G} \mathbf{e}_{k,L}
\end{aligned}
\end{equation}
where $\mathcal{I}_{n_f}$ is defined in (\ref{eq:f_hat_tao_aa}). It can be seen that $\mathbf{\bar E}_{\mathrm{id}}$ appears as multiplicative uncertainty coupled with the true augmented fault signal $\mathbf{f}_{k-\tau,L-\tau}^\zeta$ and the online I/O data $\mathbf{z}_{k,L}$.

We regard $\mathbf{f}_{k-\tau,L-\tau}^\zeta$ and $\mathbf{z}_{k,L}$ as unknown but energy bounded.
Hence $\mathbf{f}_{k-\tau,L-\tau}^\zeta$ and $\mathbf{z}_{k,L}$ in the first two terms of (\ref{eq:fest_err}) lead to an estimation bias, while the online innovation signal $\mathbf{e}_{k,L}$ in the third term causes zero mean, stochastic estimation errors. We would like to reduce the estimation bias by minimizing the matrix 2-norms $\left\| {\mathcal{T}}_{s} \left( \mathcal{G} \right) \right\|_2$ ($s=f,z$), and at the same time
minimize the Frobenius norm $\mathrm{tr} \left( \mathcal{G} \Sigma_{e,L} \mathcal{G}^\mathrm{T} \right)$ by using the available innovation covariance $\Sigma_{e,L}$. These three objectives are formulated by the following mixed-norm problem:
\begin{equation}\label{eq:offline_mixed_prob_symb}
\begin{array}{c}
\mathcal{G}_{\text{r,off}} = \argmin\limits_{\mathcal{G}}\; \mathrm{tr} \left( \mathcal{G} \Sigma_{e,L} \mathcal{G}^\mathrm{T} \right) \\
\mathrm{s.t.}\; \mathbb{\bar E} \left( {\mathcal{T}}_{s} \left( \mathcal{G} \right) {\mathcal{T}}_{s}^\text{T} \left( \mathcal{G} \right) \right) \leq \gamma_s^2 I, \; s=f, z \\
\end{array}
\end{equation}
where the matrix $\mathcal{G}$ denotes the $\tau$-delay fault estimator (\ref{eq:G_rkL}), $\mathbb{\bar E}$ denotes mathematical expectation over the identification innovations $\mathbf{\bar E}_{\mathrm{id}}$, $\gamma_f >0$ and $\gamma_z >0$ are the user-defined parameters to achieve a trade-off between estimation error variance and bias. Note that the matrix 2-norms $\left\| {\mathcal{T}}_{s} \left( \mathcal{G} \right) \right\|_2$ ($s=f,z$) are affected by the stochastic identification innovations $\mathbf{\bar E}_{\mathrm{id}}$ according to (\ref{eq:fest_err}), hence their mathematical expectations are used in (\ref{eq:offline_mixed_prob_symb}). 
Note also that it is straightforward to prove $\mathbb{\bar E} \left( {\mathcal{T}}_{s}^\mathrm{T} \left( \mathcal{G} \right) {\mathcal{T}}_{s} \left( \mathcal{G} \right) \right) \leq \gamma_s^2 I$ holds if and only if $\mathbb{\bar E} \left( {\mathcal{T}}_{s} \left( \mathcal{G} \right) {\mathcal{T}}_{s}^\mathrm{T} \left( \mathcal{G} \right) \right) \leq \gamma_s^2 I$ in (\ref{eq:offline_mixed_prob_symb}) holds. Here we use 
$\mathbb{\bar E} \left( {\mathcal{T}}_{s} \left( \mathcal{G} \right) {\mathcal{T}}_{s}^\mathrm{T} \left( \mathcal{G} \right) \right)$ in (\ref{eq:offline_mixed_prob_symb}), because it brings a clear geometrical interpretation for parameter tuning as explained later in Section \ref{sect:tune_geometric}.
With the tedious but straightforward derivations summarized in Appendix \ref{app:computations}, the above problem (\ref{eq:offline_mixed_prob_symb}) can be explicitly written as
\begin{subequations}\label{eq:offline_mixed_prob_explicit}
\begin{equation}\label{eq:offline_mixed_prob_explicit_cost}
\mathcal{G}_{\rm{r,off}} = \argmin\limits_{\mathcal{G}}\; \mathrm{tr} \left( \mathcal{G} \Sigma_{e,L} \mathcal{G}^\mathrm{T} \right)
\end{equation}
\begin{equation}\label{eq:offline_mixed_prob_explicit_constf}
\mathrm{s.t.}\;
\left[ \begin{array}{cc}
         \mathcal{G} & \mathcal{I}_{n_f}
       \end{array} \right]
\left[ \begin{array}{cc}
         \Pi_f & -\hat {\Upsilon}_{L,\tau} \\
         -\hat {\Upsilon}_{L,\tau}^\mathrm{T} & I_{n_f}
       \end{array}
 \right]
\left[ \begin{array}{c}
         \mathcal{G}^\mathrm{T} \\
         \mathcal{I}_{n_f}^\mathrm{T}
       \end{array}
 \right] \leq \gamma_f^2 I
\end{equation}
\begin{equation}\label{eq:offline_mixed_prob_explicit_constz}
\mathcal{G} \Pi_z \mathcal{G}^\mathrm{T} \leq \gamma_z^2 I,
\end{equation}
\end{subequations}
with $\Pi_f$ and $\Pi_z$ defined in (\ref{eq:pif}) and (\ref{eq:piz}), respectively.
The mixed-norm problem (\ref{eq:offline_mixed_prob_explicit}) can be easily transformed into an equivalent semi-definite programming (SDP) problem that can be solved efficiently \cite{Boyd2004}.
Since the optimization problem (\ref{eq:offline_mixed_prob_explicit}) is determined only by the identification data and does not involve any online I/O data, it can be solved offline to obtain the robust fault estimator denoted as $\mathcal{G}_{\rm{r,off}}$.

\subsection{Parameter tuning using geometric interpretation}
\label{sect:tune_geometric}
Next, we will provide a systematic method to tune the two user-defined parameters $\gamma_f^2$ and $\gamma_z^2$ by using a geometric interpretation of the mixed-norm problem (\ref{eq:offline_mixed_prob_explicit}).

\begin{figure}[!ht]
\begin{center}
\includegraphics[width=2.5in]{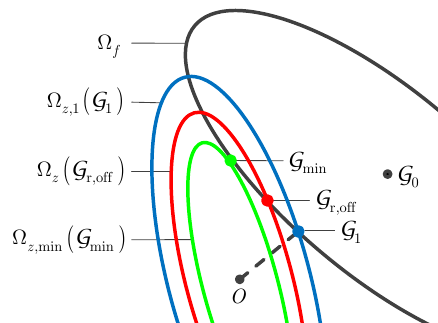}
\caption{Geometric interpretation of the mixed-norm problem (\ref{eq:offline_mixed_prob_explicit}): the constraints (\ref{eq:offline_mixed_prob_explicit_constf}) and (\ref{eq:offline_mixed_prob_explicit_constz}) define the ellipsoid $\Omega_f$ centered at $\mathcal{G}_0$ and the ellipsoid $\Omega_z$ centered at the origin $O$, respectively.
Lying at the boundary of the ellipsoid $\Omega_f$, the optimal solution
$\mathcal{G}_{\rm{r,off}}$ gives the shortest distance measured by the objective function (\ref{eq:offline_mixed_prob_explicit_cost}) from the origin to the intersection set $\Omega_f \bigcap \Omega_z$.
With $\gamma_z^2 = \gamma_{z,\rm{min}}^2$, the ellipsoid $\Omega_z$ becomes $\Omega_{z,\rm{min}} \left( \mathcal{G}_{\rm{min}} \right)$ in green which intersects with the ellipsoid $\Omega_f$ at a single point $\mathcal{G}_{\rm{min}}$.
At the boundary of the ellipsoid $\Omega_f$,
$\mathcal{G}_1$ gives the shortest distance from the origin to the ellipsoid $\Omega_f$.
The ellipsoids $\Omega_{z,1} \left( \mathcal{G}_1 \right)$ in blue and $\Omega_{z} \left( \mathcal{G}_{\rm{r,off}} \right)$ in red represent the ellipsoids $\Omega_z$ with $\mathcal{G}_1$ and $\mathcal{G}_{\rm{r,off}}$ lying at the boundary, respectively.}
\label{fig:geom}
\end{center}
\end{figure}

With some matrix manipulations,
we can see that the constraints (\ref{eq:offline_mixed_prob_explicit_constf}) and (\ref{eq:offline_mixed_prob_explicit_constz}) define two ellipsoids
\begin{equation}\label{eq:elipsoid_f}
\Omega_f = \left\{
\mathcal{G} \left| \left( \mathcal{G} - \mathcal{G}_0 \right) \Pi_f \left( \mathcal{G} - \mathcal{G}_0 \right)^\mathrm{T}
\leq \mathcal{G}_0 \Pi_f \mathcal{G}_0^\mathrm{T} - I + \gamma_f^2 I \right.
\right\},
\end{equation}
\begin{equation}
\Omega_z = \left\{ \mathcal{G} \left| \mathcal{G} \Pi_z \mathcal{G}^\mathrm{T} \leq \gamma_z^2 I \right. \right\},
\end{equation}
respectively, with $\mathcal{G}_0 = \mathcal{I}_{n_f} \hat {\Upsilon}_{L,\tau}^\mathrm{T} \Pi_f^{-1}$.
Since the objective function (\ref{eq:offline_mixed_prob_explicit_cost})
can be regarded as a measure of the distance from $\mathcal{G}$ to
the origin $0_{n_f \times \left(n_y \cdot L \right)}$, the optimization problem (\ref{eq:offline_mixed_prob_explicit}) is equivalent to finding the point $\mathcal{G}_{\rm{r,off}}$ in the set
$\Omega_f \bigcap \Omega_z$ that is closest to the origin, as shown in Fig. \ref{fig:geom}.

First, we would like to find the region of $\gamma_f^2$ and $\gamma_z^2$ so that the optimization problem (\ref{eq:offline_mixed_prob_explicit}) is feasible and non-trivial. In the case that the origin $0_{n_f \times \left(n_y \cdot L \right)} \in \Omega_f \bigcap \Omega_z$, we would have the trivial solution $\mathcal{G}_{\rm{r,off}} = 0_{n_f \times \left(n_y \cdot L \right)}$ which makes no sense for fault estimation. Hence
$0_{n_f \times \left(n_y \cdot L \right)} \notin \Omega_f$ and $\Omega_f \neq \varnothing$ are both required, which implies the region of $\gamma_f^2$ as below according to (\ref{eq:elipsoid_f}):
\begin{equation}\label{eq:gammaf_range}
1- \lambda_{\mathrm{min}} \left( \mathcal{G}_0 \Pi_f \mathcal{G}_0^\mathrm{T} \right) = \gamma_{f,\text{min}}^2 \leq \gamma_f^2 < 1.
\end{equation}

For a given $\gamma_f^2$ satisfying (\ref{eq:gammaf_range}), we solve the following optimization problem
\begin{equation}\label{eq:offline_prob_gammazmin}
\begin{array}{c}
\left\{ \mathcal{G}_{\rm{min}}, \gamma_{z,\rm{min}}^2 \right\} = \argmin\limits_{\mathcal{G}, \gamma_z^2}\; \gamma_z^2 \\
\mathrm{s.t.}\; (\ref{eq:offline_mixed_prob_explicit_constf}) \text{ and } (\ref{eq:offline_mixed_prob_explicit_constz}) \\
\end{array}
\end{equation}
whose solution gives the minimal $\gamma_z^2$, referred to as $\gamma_{z,\text{min}}^2$, that ensures $\Omega_f \bigcap \Omega_z \neq \varnothing$. Therefore, we should select $\gamma_z^2 \in \left[ \gamma_{z,\text{min}}^2, \infty \right)$ to ensure feasibility of the optimization problem (\ref{eq:offline_mixed_prob_explicit}).
The ellipsoid $\Omega_{z,\text{min}}$ in Fig. \ref{fig:geom} represents the ellipsoid $\Omega_z$ with $\gamma_z^2 = \gamma_{z,\text{min}}^2$, and its intersection with the ellipsoid $\Omega_f$ includes only the single point $\mathcal{G}_{\text{min}}$.

By discarding the constraint (\ref{eq:offline_mixed_prob_explicit_constz}) from the problem (\ref{eq:offline_mixed_prob_explicit}) and fixing $\gamma_f^2$ at the same given value as in (\ref{eq:offline_prob_gammazmin}), we formulate another problem
\begin{equation}\label{eq:offline_prob_gammazmax}
\begin{array}{c}
\mathcal{G}_1 = \argmin\limits_{\mathcal{G}}\; \mathrm{tr} \left( \mathcal{G} \Sigma_{e,L} \mathcal{G}^\mathrm{T} \right) \\
\mathrm{s.t.}\; (\ref{eq:offline_mixed_prob_explicit_constf}) 
\end{array}
\end{equation}
Because the optimal solution $\mathcal{G}_1$ gives the shortest distance from the origin to the ellipsoid $\Omega_f$, and moreover $0_{n_f \times \left(n_y \cdot L \right)} \notin \Omega_f$, the solution $\mathcal{G}_1$ must lie at the boundary of the ellipsoid $\Omega_f$, as shown in Fig. \ref{fig:geom}.
Define $\gamma_{z,1}^2 = \lambda_{\text{max}} \left(\mathbb{\bar E} \left( {\mathcal{T}}_{z} \left( \mathcal{G}_1 \right) {\mathcal{T}}_{z}^\mathrm{T} \left( \mathcal{G}_1 \right) \right)\right)$. Let the ellipsoid $\Omega_{z,1}$ in Fig. \ref{fig:geom} represent the set $\Omega_z$ with $\gamma_z^2 = \gamma_{z,1}^2$, and it has the solution  $\mathcal{G}_1$ at its boundary.

\begin{table*}[!t]
 \caption{Trade-offs between fault estimation bias and error variance of the robust fault estimator $\mathcal{G}_{\rm{r,off}}$ at time instant $k$ when tuning user-defined parameters $\gamma_f^2$ and $\gamma_z^2$ in (\ref{eq:offline_mixed_prob_explicit}): ``Constant'', ``$\nearrow$'', and ``$\searrow$'' means that the performance criterion in the corresponding column remains constant, monotonically increases, and monotonically decreases with regard to the user-defined parameter specified in the corresponding row, respectively.}
 \label{tab:perf_tradeoff}
 \centering
 \begin{tabular}{lccc}
  \toprule
   User-defined & First bias term & Second bias term & Variance \\
   parameters & $\mathbb{\bar E} \left\|  {\mathcal{T}}_{f} \left( \mathcal{G}_{\text{r,off}} \right) 
   	\mathbf{f}_{k-\tau, L-\tau}^\zeta \right\|_2^2 $
   & $\mathbb{\bar E} \left\|  {\mathcal{T}}_{z} \left( \mathcal{G}_{\text{r,off}} \right) 
   	\mathbf{z}_{k, L} \right\|_2^2 $
   & $\mathrm{tr} \left( \mathcal{G}_{\rm{r,off}} \Sigma_{e,L} \mathcal{G}_{\rm{r,off}}^\mathrm{T} \right)$\\
  \midrule
  $\gamma_z^2 \in \left[ \gamma_{z,\rm{min}}^2, \gamma_{z,1}^2 \right]$ & Constant & $\nearrow$ & $\searrow$ \\
  $\gamma_z^2 \in \left[ \gamma_{z,1}^2, \infty \right)$ & Constant & Constant & Constant \\
  $\gamma_f^2 \in \left[ \gamma_{f,\rm{min}}^2, 1 \right)$ & $\nearrow$ & $\searrow$ & $\searrow$ \\
  \bottomrule
 \end{tabular}
\end{table*}

Similarly to the above obtained solution $\mathcal{G}_1$ of the problem (\ref{eq:offline_prob_gammazmax}), the solution $\mathcal{G}_{\rm{r,off}}$ of the problem (\ref{eq:offline_mixed_prob_explicit}) also lies at the boundary of the ellipsoid $\Omega_f$. This allows the three terms of the fault estimation error in (\ref{eq:fest_err}) to be explained using Fig. \ref{fig:geom}:
\begin{enumerate}
  \item[1)] The bias related to the first term $\mathcal{T}_f \left( \mathcal{G} \right) \mathbf{f}_{k-\tau, L-\tau}^\zeta$ is determined by the size of the ellipsoid $\Omega_f$;
  \item[2)] The bias related to the second term $\mathcal{T}_z \left( \mathcal{G} \right) \mathbf{z}_{k,L}$ is determined by the size of the ellipsoid $\Omega_z \left( \mathcal{G}_{\rm{r,off}} \right)$ with $\mathcal{G}_{\rm{r,off}}$ lying on its boundary, i.e., the ellipsoid $\Omega_z$ with $\gamma_z^2 = \lambda_{\text{max}} \left(\mathbb{\bar E} \left( {\mathcal{T}}_{z} \left( \mathcal{G}_{\text{r,off}} \right) {\mathcal{T}}_{z}^\mathrm{T} \left( \mathcal{G}_{\text{r,off}} \right) \right)\right)$;
  \item[3)] The fault estimation error variance related to the third term $\mathcal{G} \mathbf{e}_{k,L}$  is represented by the distance from the origin to the optimal solution $\mathcal{G}_{\rm{r,off}}$.
\end{enumerate}

With the above basic geometric interpretation, we can analyze the performance trade-offs of the robust fault estimator $\mathcal{G}_{\rm{r,off}}$ when tuning $\gamma_f^2 \in \left[ \gamma_{f,\text{min}}^2, 1 \right)$ and $\gamma_z^2 \in \left[ \gamma_{z,\text{min}}^2, \infty \right)$, as shown in Table \ref{tab:perf_tradeoff}.
First, we fix $\gamma_f^2$ and tune $\gamma_z^2$. In this case, the ellipsoid $\Omega_f$ is fixed, which makes the first bias term in the first two rows of Table \ref{tab:perf_tradeoff} remain constant.
With the fixed $\gamma_f^2$, by increasing $\gamma_{z}^2$ from $\gamma_{z,\rm{min}}^2$ towards $\gamma_{z,1}^2$, the intersection set $\Omega_f \bigcap \Omega_z$ becomes larger, and the optimal solution $\mathcal{G}_{\rm{r,off}}$ moves from the point $\mathcal{G}_{\rm{min}}$ along the boundary of the ellipsoid $\Omega_f$ towards the point $\mathcal{G}_1$.
When we further increase $\gamma_z^2$ for $\gamma_z^2 \geq \gamma_{z,1}^2$, the optimal solution $\mathcal{G}_{\rm{r,off}}$ of the problem (\ref{eq:offline_mixed_prob_explicit}) would remain located at the point $\mathcal{G}_{1}$, because $\mathcal{G}_{1}$ satisfies both constraints (\ref{eq:offline_mixed_prob_explicit_constf}) and
(\ref{eq:offline_mixed_prob_explicit_constz}) and gives the shortest distance to the origin according to the problem (\ref{eq:offline_prob_gammazmax}).
Therefore, the size of the ellipsoid $\Omega_z \left( \mathcal{G}_{\rm{r,off}} \right)$, which determines the second estimation bias term in the first two rows of Table \ref{tab:perf_tradeoff}, monotonically increases for
$\gamma_z^2 \in \left[ \gamma_{z,\rm{min}}^2, \gamma_{z,1}^2 \right)$ and remains constant for
$\gamma_z^2 \in \left[ \gamma_{z,1}^2, \infty \right)$. The distance from the origin to $\mathcal{G}_{\rm{r,off}}$, which determines the fault estimation error variance in the first two rows of Table \ref{tab:perf_tradeoff}, monotonically decreases for
$\gamma_z^2 \in \left[ \gamma_{z,\rm{min}}^2, \gamma_{z,1}^2 \right)$ and remains constant for $\gamma_z^2 \in \left[ \gamma_{z,1}^2, \infty \right)$. For the third row of Table \ref{tab:perf_tradeoff}, we tune $\gamma_f^2$ and select a sufficiently large value of $\gamma_z^2$ that ensures the problem (\ref{eq:offline_mixed_prob_explicit}) to be feasible. With $\gamma_f^2$ increasing, the size of the ellipsoid $\Omega_f$, which determines the first bias term in the third row of Table \ref{tab:perf_tradeoff}, monotonically increases. Meanwhile, the optimal solution
$\mathcal{G}_{\rm{r,off}}$, which lies at the boundary of the ellipsoid $\Omega_f$, moves closer to the origin. Therefore, both the second bias term and the fault estimation error variance in the third row of Table \ref{tab:perf_tradeoff}, which are determined by the size of the ellipsoid $\Omega_z \left( \mathcal{G}_{\rm{r,off}} \right)$ and the distance from the origin to the point $\mathcal{G}_{\rm{r,off}}$, monotonically decrease.

We summarize the data-driven robust design in Algorithm \ref{alg:offline_robust_design}. The nominal design $\mathcal{G}_{\rm{n}}$ obtained from Algorithm \ref{alg:nominal_design} can be used as a benchmark for tuning $\gamma_f^2$ and $\gamma_z^2$ in Step 2 of Algorithm \ref{alg:offline_robust_design}. For example, compared to the nominal design, the robust design achieves smaller averaged worst-case bias if $\gamma_s^2 \leq \lambda_{\text{max}} \left(\mathbb{\bar E} \left( {\mathcal{T}}_{s} \left( \mathcal{G}_\text{n} \right) {\mathcal{T}}_{s}^\mathrm{T} \left( \mathcal{G}_\text{n} \right) \right)\right)$ ($s=f,z$).

\begin{algorithm}
  \caption{Data-driven robust RH fault estimation}
  \label{alg:offline_robust_design}
  \begin{algorithmic}
    \State
    \begin{enumerate}
      \item[1)] Complete the steps 1-3 in Algorithm \ref{alg:nominal_design}; compute $M_i^u$, $M_i^y$, and $M_i^f$ according to (\ref{eq:Muy}) and (\ref{eq:Mif}).
      \item[2)] Tune $\gamma_f^2 \in \left[ \gamma_{f,\rm{min}}^2, 1 \right)$ and $\gamma_z^2 \in \left[ \gamma_{z,\rm{min}}^2, \infty \right)$ according to the performance trade-offs shown in Table \ref{tab:perf_tradeoff}, where $\gamma_{f,\rm{min}}^2$ and $\gamma_{z,\rm{min}}^2$ are obtained from the optimization problems (\ref{eq:gammaf_range}) and (\ref{eq:offline_prob_gammazmin}) respectively.
      \item[3)] Solve the problem (\ref{eq:offline_mixed_prob_explicit}) to compute the robust RH fault estimator $\mathcal{G}_{\mathrm{r,off}}$.
    \end{enumerate}
  \end{algorithmic}
\end{algorithm}

\section{Data-driven robust receding horizon fault estimation with online optimization}\label{sect:dd_robust_onlineopt}
The online I/O data is regarded as unknown in Algorithm \ref{alg:offline_robust_design}. In order to better exploit the available online data, this section proposes an online mixed-norm optimization approach. This can further reduce the estimation errors when the online I/O data have large amplitudes, at the expense of increased computational burden.

\subsection{Online mixed-norm problem}
With the notation
\begin{equation}\label{eq:bar_beta}
\bar \beta_{k,L} = \mathbf{\bar M}_L^z \mathbf{z}_{k,L},
\end{equation}
we divide $\bar \beta_{k,L}$ into $L$ row blocks as in
\begin{equation}\label{eq:bar_beta_i}
\bar \beta_{k,L} = \left[ \begin{array}{cccc}
                            \beta_{k,1}^\mathrm{T} & \beta_{k,2}^\mathrm{T} & \cdots & \beta_{k,L}^\mathrm{T}
                          \end{array}
 \right]^\mathrm{T},
\end{equation}
with $\beta_{k,i} \in \mathbb{R}^{N}$. Then the term $\mathcal{G} \mathbf{\bar E}_{\mathrm{id}} \mathbf{\bar M}_L^z \mathbf{z}_{k,L}$ in (\ref{eq:fest_err}) can be rewritten as
\begin{equation}\label{eq:GEMz}
\begin{aligned}
&\,\mathcal{G} \mathbf{\bar E}_{\mathrm{id}} \mathbf{\bar M}_L^z \mathbf{z}_{k,L}
= \mathcal{G} \mathbf{\bar E}_{\mathrm{id}} \bar \beta_{k,L} \\
=&\, \mathcal{G}
\left[ \begin{array}{c}
         \mathbf{E}_{\mathrm{id}} \beta_{k,1} \\
         \mathbf{E}_{\mathrm{id}} \beta_{k,2} \\
         \vdots \\
         \mathbf{E}_{\mathrm{id}} \beta_{k,L}
       \end{array} \right]
= \mathcal{G}
\underbrace{\left[ \begin{array}{c}
         \beta_{k,1}^\mathrm{T} \otimes I_{n_y} \\
         \beta_{k,2}^\mathrm{T} \otimes I_{n_y} \\
         \vdots \\
         \beta_{k,L}^\mathrm{T} \otimes I_{n_y}
       \end{array} \right]}_{{\Gamma_{k,L}}}
  \mathrm{vec}\left( \mathbf{E}_{\mathrm{id}} \right)
\end{aligned}
\end{equation}
according to the property of Kronecker product \cite{Brew1978}. Based on (\ref{eq:GEMz}), the estimation error in (\ref{eq:fest_err}) becomes
\begin{equation}
\Delta f(k-\tau) = \mathcal{T}_{f} \left( \mathcal{G} \right) \mathbf{f}_{k-\tau,L-\tau}^\zeta - \mathcal{G} \Gamma_{k,L}  \mathrm{vec}\left( \mathbf{E}_{\mathrm{id}} \right) + \mathcal{G} \mathbf{e}_{k,L}.
\end{equation}
Then the statistics of $\mathrm{vec} \left( \mathbf{E}_{\mathrm{id}} \right)$, i.e.,
$$\mathbb{E} \left( \mathrm{vec} \left( \mathbf{E}_{\mathrm{id}} \right)  \mathrm{vec} \left( \mathbf{E}_{\mathrm{id}} \right)^\mathrm{T} \right)
= I_N \otimes \Sigma_e,
$$ can be exploited to evaluate the fault estimation error variance.
Therefore, we formulate the following optimization problem similarly to (\ref{eq:offline_mixed_prob_symb}):
\begin{equation}\label{eq:online_mixed_prob_symb}
\begin{array}{c}
\mathcal{G}_{\rm{r,on}} = \argmin\limits_{\mathcal{G}}\; \mathrm{tr}
\left( \mathcal{G} \Sigma_{e,L} \mathcal{G}^\mathrm{T}
 + \mathcal{G} \Gamma_{k,L} \left( I_N \otimes \Sigma_e \right)
 \Gamma_{k,L}^\mathrm{T} \mathcal{G}^\mathrm{T} \right) \\
\mathrm{s.t.}\;\; \mathbb{\bar E} \left( {\mathcal{T}}_{f} \left( \mathcal{G} \right) {\mathcal{T}}_{f}^\text{T} \left( \mathcal{G} \right) \right) \leq \gamma_f^2 I
\end{array}
\end{equation}
with the user-defined parameter $\gamma_f$. The constraint in the above optimization problem (\ref{eq:online_mixed_prob_symb}) can be explicitly written as (\ref{eq:offline_mixed_prob_explicit_constf}).
The optimization problem (\ref{eq:online_mixed_prob_symb}) has to be solved at each time instant to update the robust fault estimator $\mathcal{G}_{\rm{r,on}}$ because $\Gamma_{k,L}$ in the cost function is determined by the online I/O data according to (\ref{eq:bar_beta})-(\ref{eq:GEMz}).

\subsection{Parameter tuning using geometric interpretation}
Since the online mixed-norm problem (\ref{eq:online_mixed_prob_symb}) has the structure similar to that of the offline mixed-norm problem (\ref{eq:offline_mixed_prob_explicit}), the performance trade-offs by tuning $\gamma_f$ in (\ref{eq:online_mixed_prob_symb}) are also similar to that explained in Table \ref{tab:perf_tradeoff}.


The proposed data-driven robust fault estimation with online optimization is summarized in Algorithm \ref{alg:online_robust_design}.
In order to reduce the computational burden of online optimization, the problem (\ref{eq:online_mixed_prob_symb}) is implemented only if the estimation bias of the offline designed fault estimator is larger than a user-defined threshold $\alpha$, as shown in Step 2 of Algorithm \ref{alg:online_robust_design}.


The offline designed fault estimator $\mathcal{G}_{\mathrm{r,off}}$ from Algorithm \ref{alg:offline_robust_design} can be used as a benchmark for tuning $\gamma_f^2$ in Step 2.2 of Algorithm \ref{alg:online_robust_design}. For example, compared to $\mathcal{G}_{\mathrm{r,off}}$, the online optimization (\ref{eq:online_mixed_prob_symb}) achieves smaller averaged worst-case bias if $\gamma_f^2 \leq \lambda_{\text{max}}\left(\mathbb{\bar E} \left( {\mathcal{T}}_{f} \left( \mathcal{G}_{\text{r,off}} \right) {\mathcal{T}}_{f}^\text{T} \left( \mathcal{G}_{\text{r,off}} \right) \right)\right)$.

\begin{algorithm}
  \caption{Data-driven robust RH fault estimation with online optimization}
  \label{alg:online_robust_design}
  \begin{algorithmic}
    \State
    \begin{enumerate}
      \item[1)] Follow Algorithm \ref{alg:offline_robust_design} to compute the offline designed fault estimator $\mathcal{G}_{\mathrm{r,off}}$.
      \item[2)] If $\lambda_\text{min} \left( \mathbb{\bar E} \left( \mathcal{T}_{z} \left(  \mathcal{G}_{\mathrm{r,off}} \right) \mathcal{T}_{z}^\mathrm{T} \left(  \mathcal{G}_{\mathrm{r,off}} \right) \right) \right)  \left\| \mathbf{z}_{k,L} \right\|_2^2 > \alpha$ ($\alpha$ is a user-defined threshold), the online optimization in the following steps is implemented; otherwise, the offline designed estimator $\mathcal{G}_{\mathrm{r,off}}$ is used.
          \begin{enumerate}
            \item[2.1)] Compute $\Gamma_{k,L}$ according to (\ref{eq:bar_beta})-(\ref{eq:GEMz}).
            \item[2.2)] Tune $\gamma_f^2 \in \left[ \gamma_{f,\text{min}}^2, 1 \right)$ similarly to Step 2 of Algorithm \ref{alg:offline_robust_design}, with $\gamma_{f,\text{min}}^2$ defined in (\ref{eq:gammaf_range}).
            \item[2.3)] Solve the problem (\ref{eq:online_mixed_prob_symb}) to compute the robust RH fault estimator $\mathcal{G}_{\mathrm{r,on}}$.
          \end{enumerate}
    \end{enumerate}
  \end{algorithmic}
\end{algorithm}


\section{Simulation studies}\label{sect:sim}
Consider a continuous-time linearized vertical take-off and landing (VTOL) aircraft model that has been studied in \cite{Dong2012a, Dong2012b, Dong2012c, Gust2001}:
\begin{align*}
{\dot x}_c (t) &= A_c x_c(t) + B_c u_c(t), \\
y_c(t) &= C_c (t), \\
A_c &= \left[ \begin{smallmatrix}
                -0.0366 & 0.0271 & 0.0188 & -0.4555 \\
                0.0482 & -1.01 & 0.0024 & -4.0208 \\
                0.1002 & 0.3681 & -0.707 & 1.42 \\
                0 & 0 & 1 & 0
              \end{smallmatrix}
 \right], \\
B_c &= \left[ \begin{smallmatrix}
                0.4422 & 0.1761 \\
                3.5446 & -7.5922 \\
                -5.52 & 4.49 \\
                0 & 0
              \end{smallmatrix}
 \right],\;
C_c = \left[ \begin{smallmatrix}
               1 & 0 & 0 & 0 \\
               0 & 1 & 0 & 0 \\
               0 & 0 & 1 & 0 \\
               0 & 1 & 1 & 1
             \end{smallmatrix}
 \right].
\end{align*}
With a sampling rate of 0.5 seconds, the discrete-time model (\ref{eq:sys}) is obtained, with $D=0$ and $F=I_4$. The process and measurement noises, $w(k)$ and $v(k)$, are assumed to be zero mean white noises, respectively with covariances of $Q = 0.16 I_4$ and $R = 0.64 I_4$.

Since the open-loop plant is unstable, an empirical stabilizing output feedback controller is used \cite{Dong2012c}, i.e.,
\begin{equation}\label{eq:controller}
u(k) = - \left[ \begin{smallmatrix}
                  0 & 0 & -0.5 & 0 \\
                  0 & 0 & -0.1 & -0.1
                \end{smallmatrix}
 \right] y(k) + \eta(k),
\end{equation}
where $\eta(k)$ is the reference signal.

In the identification experiment, the reference signal $\eta(k)$ is zero-mean white noise with the covariance of $\mathrm{diag}\left( 1,1 \right)$, which ensures persistent excitation.
We collect $N=1000$ data samples from the identification experiment. In the identification algorithm, the past horizon is selected as $p=10$.

The considered fault cases include:
\begin{itemize}
  \item Actuator faults: $E=B$, $G=D$,
  \item Sensor faults: $E = 0_{4 \times 2}$, $G= \left[ \begin{smallmatrix}
                                              1 & 0 & 0 & 0 \\
                                              0 & 1 & 0 & 0
                                            \end{smallmatrix} \right]^\mathrm{T}$.
\end{itemize}
The case of simultaneous actuator and sensor faults is not included here, because all the considered algorithms can be applied to the simultaneous scenario in a straightforward way, and their performance comparisons are the same as in the case of separate actuator or sensor faults.

The simulated fault signals in both fault cases are the same:
$$ f(k) = \left\{ \begin{array}{ll}
   \left[ \begin{array}{cc}
            0 & 0
           \end{array}
   \right]^\mathrm{T}, & 0 \leq k \leq 50, \\
   \left[ \begin{array}{cc}
            \mathrm{sin}\left( 0.1 \pi k \right) & 1
           \end{array}
   \right]^\mathrm{T}, & k > 50.
   \end{array}
   \right. $$

We will compare the following fault estimation methods:
\begin{itemize}
  \item Alg0: the RH fault estimator using accurate Markov parameters, described in Section \ref{sect:RHFE_dd_nominal}.
  \item DONG: the method proposed by \cite{Dong2012c}.
  \item Alg1: the data-driven nominal RH fault estimator $\mathcal{G}_{\rm{n}}$ proposed in Algorithm \ref{alg:nominal_design};
  \item Alg2: the data-driven robust RH fault estimator $\mathcal{G}_{\mathrm{r,off}}$ proposed in Algorithm \ref{alg:offline_robust_design}; in Step 3 of Algorithm \ref{alg:offline_robust_design}, we select $\gamma_f^2 = \lambda_{\text{max}} \left( \mathbb{\bar E} \left( {\mathcal{T}}_{f} \left( \mathcal{G}_{\text{n}} \right) {\mathcal{T}}_{f}^\text{T} \left( \mathcal{G}_{\text{n}} \right) \right) \right)$, and
      \begin{equation}\label{eq:alg2_gammaz}
      \gamma_z^2 = 0.5 \left( \gamma_{z,\text{min}}^2 + \gamma_{z,1}^2 \right).
      \end{equation}
  \item Alg3: the data-driven robust RH fault estimator $\mathcal{G}_{\mathrm{r,on}}$ with online optimization, proposed in Algorithm \ref{alg:online_robust_design}; in Step 2 of Algorithm \ref{alg:online_robust_design}, we select $\alpha=300$ as the threshold to determine whether or not the online optimization should be implemented; $\gamma_f^2$ is set to the same value as in Alg2.
\end{itemize}

We select the estimation horizon length $L=30$ for the considered five algorithms.

\begin{figure}[h]
	\centering
	\includegraphics[width=8.5cm]{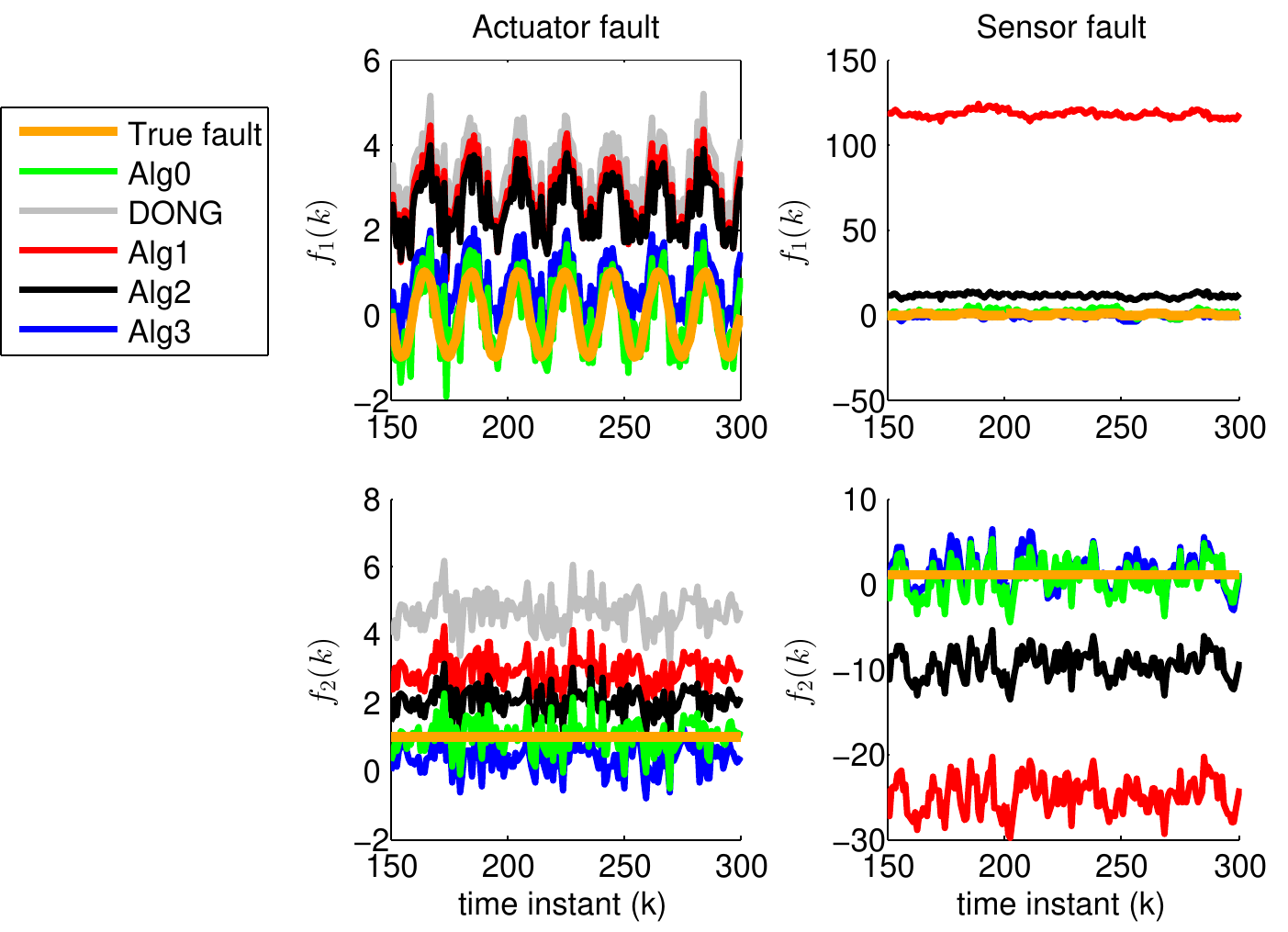}
	\caption{True fault signal and fault estimates from different algorithms.}
	\label{fig:plotfest}
\end{figure}

\begin{figure}[h]
\centering
\subfigure[Actuator faults]{
\includegraphics[width=6cm]{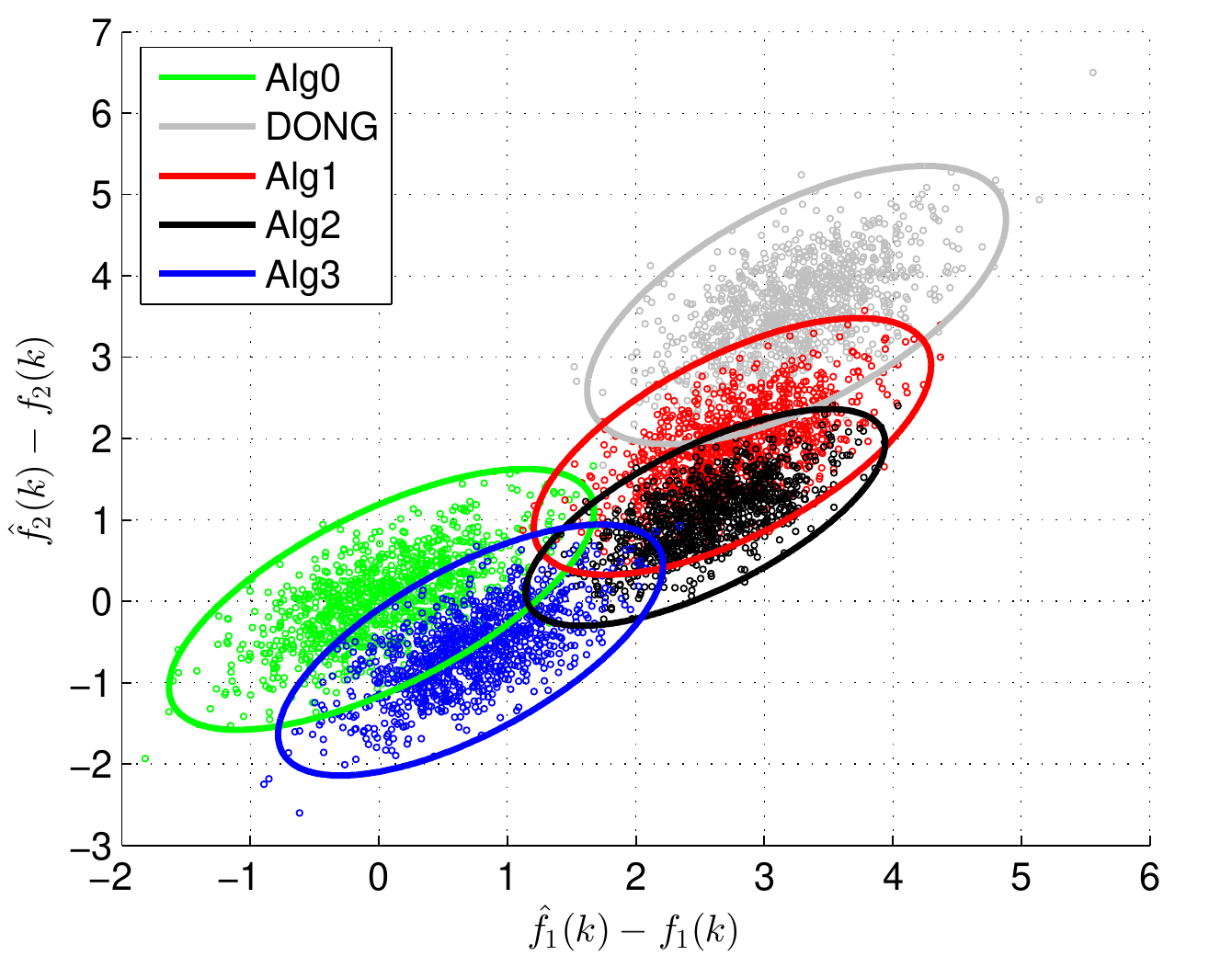}
\label{fig:result_ref1_actuator}
}
\subfigure[Sensor faults]{
\includegraphics[width=6cm]{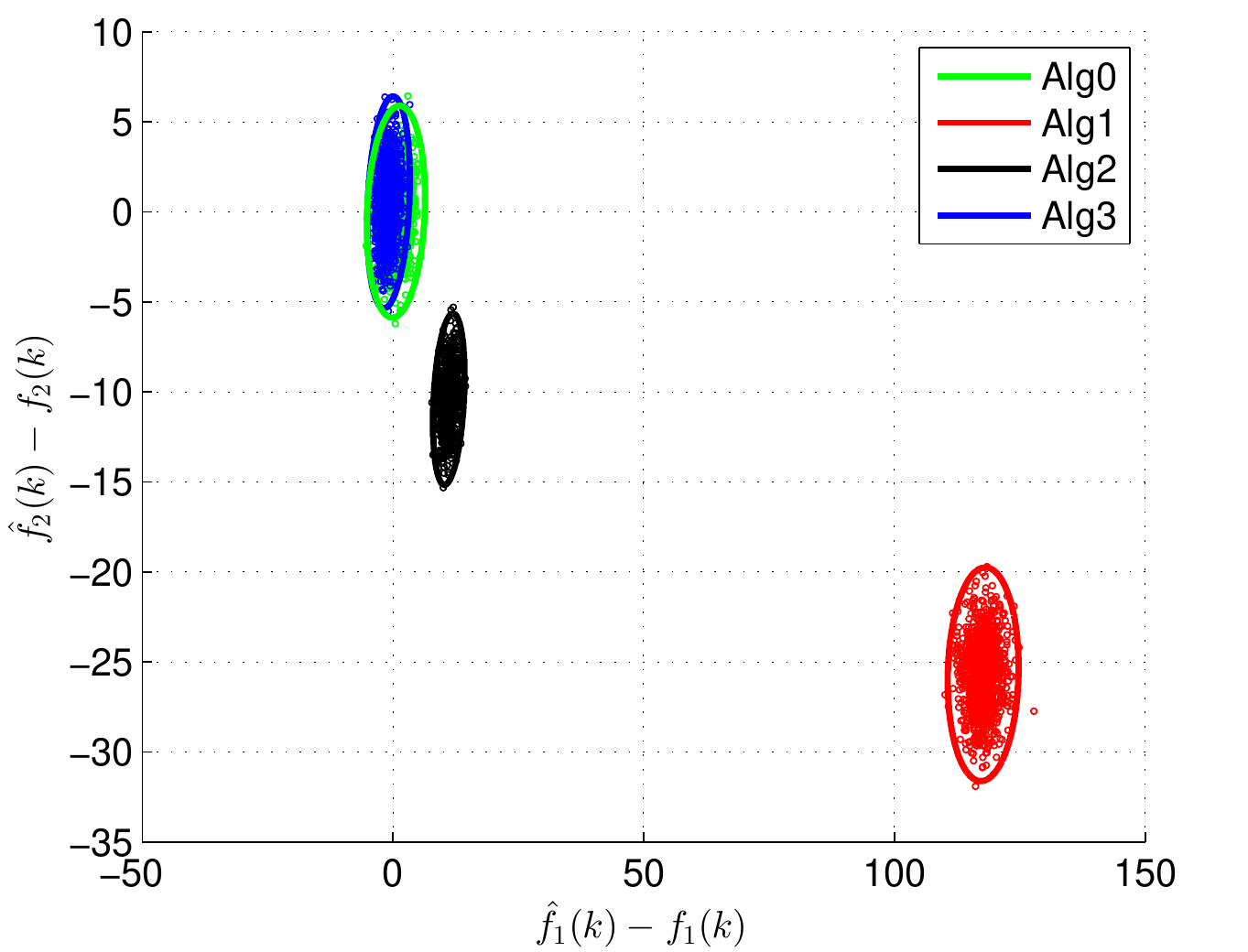}
\label{fig:result_ref1_sensor}
}
\caption{Distribution of fault estimation errors when $\eta(k)=15$. Circles: 1000 estimation errors by different fault estimation algorithms based on 1000 online I/O data samples. Ellipses: the $3\sigma$-contour of the approximated two-dimensional Gaussian distribution of the 1000 estimation errors, i.e., the contour at $\left[ \hat f(k) - f(k) \right]^\mathrm{T} {\text{cov}^{-1} \left( \hat f(k) \right)}
\left[ \hat f(k) - f(k) \right] = 3$.}
\label{fig:result_ref1}
\end{figure}

In order to show the necessity of compensating for the identification errors, we make the identification-error-effect term $\mathcal{T}_z \left( \mathcal{G} \right) \cdot \mathbf{z}_{k,L}$
in (\ref{eq:fest_err}) significantly large by setting $\eta(k) = 15$. Fault estimates from the above five algorithms are illustrated in Fig. \ref{fig:plotfest}, and the distributions of their fault estimation errors are shown in Fig. \ref{fig:result_ref1}.
By using accurate Markov parameters, Alg0 achieves unbiased fault estimation in both fault scenarios. Note that DONG cannot be directly applied to sensor faults in the unstable open-loop VTOL model \cite{Dong2012c}, hence it is not included in Fig. \ref{fig:plotfest} and \ref{fig:result_ref1_sensor} for sensor faults.
Because of neglecting the effect of identification errors, both Alg1 and DONG yield estimation biases even larger than the amplitude of true faults. In comparison, Alg2 obtains its robustness to identification error by solving an offline mixed-norm  problem, as shown in Fig. \ref{fig:result_ref1_actuator}. However, the poor performance of Alg2 in our sensor fault case (Fig. \ref{fig:result_ref1_sensor}) shows the limitation of neglecting the online availability of I/O data in the offline mixed-norm problem. 
Compared to Alg2, Alg3 significantly reduces estimation bias, as shown in Fig. \ref{fig:result_ref1_sensor}, by formulating an online mixed-norm problem to exploit online I/O data. This performance improvement is achieved at the cost of higher online computational burden. When implemented with YALMIP \cite{Lofberg2004} in the MATLAB2011b environment, on a computer with a 3.4 GHz processor and 8 GB RAM, the averaged and peak computational time per sample of Alg3 are 1.70s and 2.05s for the estimation horizon length $L=30$, while those of Alg2 are $8.37 \times 10^{-6}$s and $3.17 \times 10^{-5}$s respectively. We will investigate the computational efficiency of Alg3 for real-time implementation in future work.

\begin{figure}[h]
\centering
\includegraphics[width=7cm]{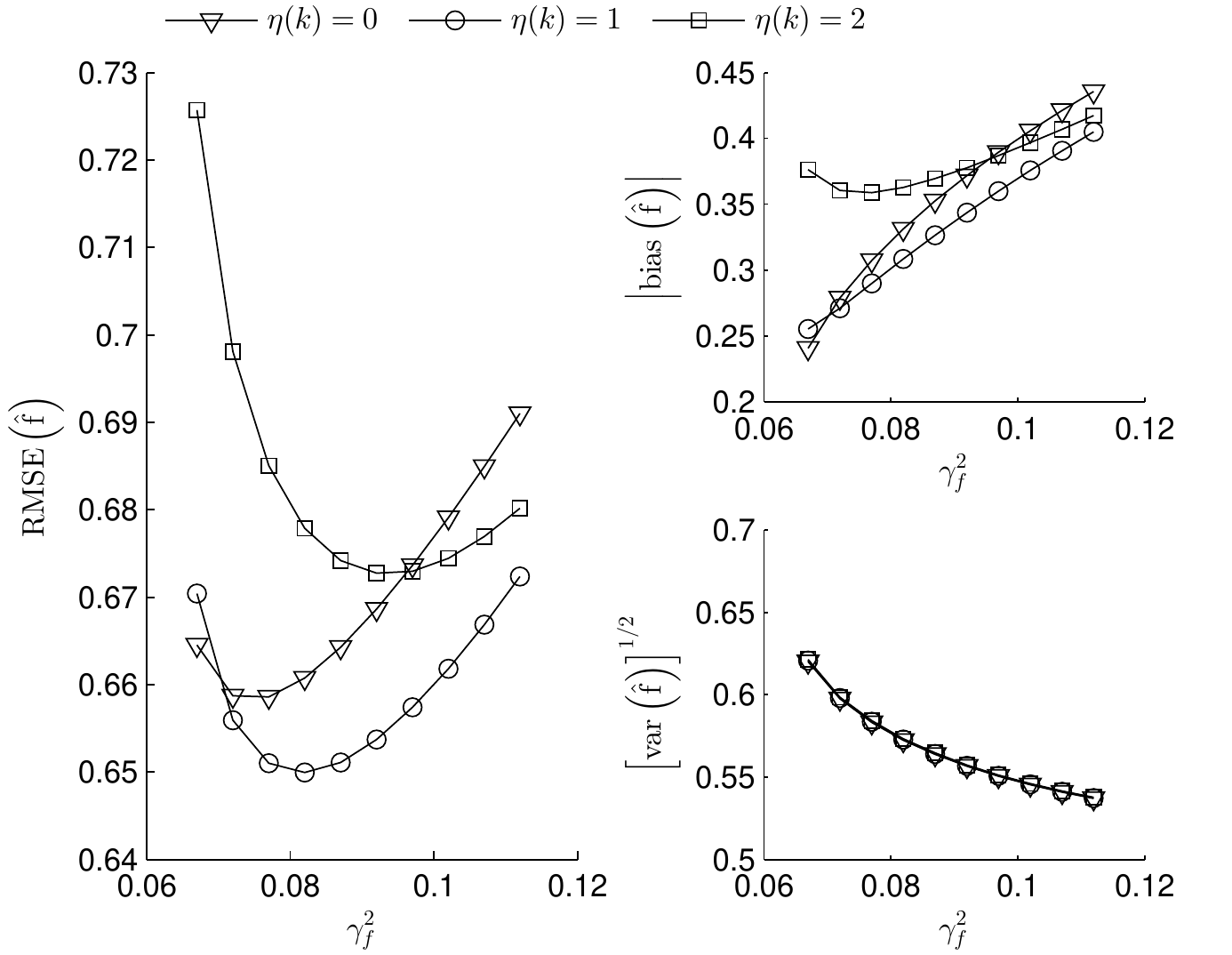}
\caption{Estimation performance of Alg2 when tuning $\gamma_f^2$ under different reference signal $\eta(k)$}
\label{fig:result_tune}
\end{figure}

In order to illustrate the performance trade-offs of Alg2, we set $\gamma_z^2$ as in (\ref{eq:alg2_gammaz}) and tune $\gamma_f^2$ under the condition of different reference signals $\eta(k)$. Fig. \ref{fig:result_tune} shows how the fault estimation bias, error variance and root mean square error (RMSE) vary with $\gamma_f^2$, which can be explained as follows using Table \ref{tab:perf_tradeoff}.
According to the fault estimation error analysis in (\ref{eq:fest_err}), the fault estimation bias is related to both $\mathcal{T}_f \left( \mathcal{G}_{\mathrm{r,off}} \right) \mathbf{f}_{k-\tau,L-\tau}^\zeta$
and $\mathcal{T}_z \left( \mathcal{G}_{\mathrm{r,off}} \right) \mathbf{z}_{k,L}$.
For $\eta(k)=0$ or $\eta(k)=1$, the online I/O data $\mathbf{z}_{k,L}$ have small amplitude, thus the total estimation bias is dominated by the bias related to  $\mathcal{T}_f \left( \mathcal{G}_{\mathrm{r,off}} \right) \mathbf{f}_{k-\tau,L-\tau}^\zeta$ which monotonically increases with $\gamma_f^2$ according to the third row of Table \ref{tab:perf_tradeoff}. This explains the fault estimation bias curves for $\eta(k)=0$ and $\eta(k)=1$ in Fig. \ref{fig:result_tune}.
For $\eta(k)=2$, the online I/O data $\mathbf{z}_{k,L}$ have relatively large amplitudes, hence for relatively small values of $\gamma_f^2$ the total estimation bias is dominated by the bias related to $\mathcal{T}_z \left( \mathcal{G}_{\mathrm{r,off}} \right) \mathbf{z}_{k,L}$ which monotonically decreases with $\gamma_f^2$, and for relatively large values of $\gamma_f^2$ the total estimation bias is dominated by the bias related to $\mathcal{T}_f \left( \mathcal{G}_{\mathrm{r,off}} \right) \mathbf{f}_{k-\tau,L-\tau}^\zeta$ which monotonically increases with $\gamma_f^2$, according to the third row of Table \ref{tab:perf_tradeoff}. This explains the fault estimation bias curve for $\eta(k)=2$ in Fig. \ref{fig:result_tune}. The monotonic decrease of the fault estimation error variances with $\gamma_f^2$ can be directly explained with the third row of Table \ref{tab:perf_tradeoff}.
As the objective function of the optimization problem (\ref{eq:offline_mixed_prob_explicit}), the fault estimation error variance $\rm{tr} \left( \mathcal{G}_{\rm{r,off}} \Sigma_{e,L} \mathcal{G}_{\rm{r,off}}^\mathrm{T} \right)$ for different reference signals $\eta(k)$ is the same because it does not depend on the reference signal $\eta(k)$. Combining the increase of fault estimation bias and the decrease of fault estimation error variance with $\gamma_f^2$, there exist the optimal $\gamma_{f,*}^2 \in \left( \gamma_{f,\text{min}}^2, 1 \right)$ such that the RMSE achieves its minimal value, as can be seen in Fig. \ref{fig:result_tune}. It is also shown that the minimal RMSE is achieved at a larger value of $\gamma_{f,*}^2$ when the amplitude of $\eta(k)$ increases, because the online I/O data have larger amplitudes with larger $\eta(k)$, thus the decrease of the bias related to $\mathcal{T}_z \left( \mathcal{G}_{\mathrm{r,off}} \right) \mathbf{z}_{k,L}$ dominates the fault estimation bias. Based on the above insights, we can anticipate how the estimation performance of Alg2 varies with different $\gamma_z^2$ for a fixed $\gamma_f^2$, as well as the performance trade-offs of Alg3. Their performance curves are not plotted due to the space limitation.

From the simulation results with different lenghts $L$ of the estimation horizon (omitted for the sake of brevity), it can be seen that the fault estimation bias and variance of Alg0, Alg1, Alg2, and Alg3 decrease with the increasing length $L$ of the estimation horizon. Straightforward proof of this observation can be derived for Alg0 using accurate Markov parameters (following Section 3.4.3 of \cite{Kailath2000}), whereas analytical proof is difficult for Alg1, Alg2, and Alg3 that rely on the identified Markov parameters contaminated with identification errors.

\section{Conclusions}
This paper has investigated data-driven fault estimation and its robustness against stochastic identification errors. First, we proposed an RH fault estimator that can be parameterized with the predictor Markov parameters. Its condition for unbiasedness generalizes that of a recently reported data-driven fault estimation method. An immediate benefit is that our proposed method can be applied to sensor faults of an unstable open-loop plant which could not be directly addressed previously. In the formulated RH fault estimator, the identification errors appear as multiplicative model uncertainty coupled with the unknown faults and the online I/O data. Then, two mixed-norm problems were formulated to enhance robustness. One can be solved offline by regarding the online I/O data as unknown signals. The other further reduces estimation errors for larger I/O data by exploiting their online availability in the mixed-norm problem, and it requires online optimization. Based on geometric interpretations of the mixed-norm problems, systematic methods were given to tune the user-defined parameters therein. 
Comparisons using a simulated aircraft model illustrated the advantages and the effectiveness of our proposed method.

\begin{ack}                               
The research leading to these results has received funding from the European Union's Seventh Framework Programme (FP7-RECONFIGURE/2007--2013) under grant agreement No. 314544.  
\end{ack}

\bibliographystyle{plain}        
\bibliography{bib_DD_RHfaultest}           

\appendix
\section{Lemmas for Theorem \ref{thm:unbias}}\label{app:property}

\begin{lem}\label{lem:zero_dyn}
Define $x_{e}(0) \in \mathbb{R}^n$, $f_{e}(i) \in \mathbb{R}^{n_f}$, and $r_{e}(i) \in \mathbb{R}^{n_y}$ ($i \geq 0$) as the initial state, input and output signal of the fault subsystem $(\Phi, \tilde E, C, G)$, respectively. 
There exists a non-zero initial state $x_{e}(0)$ such that $r_{e}(0) = r_{e}(1) = \cdots = r_{e}(L) = 0$ for all $L \geq \nu+\tau$, if and only if 
\begin{enumerate}
	\item[(\romannumeral1)] $\mathcal{O}_{\tau} x_{e}(0) =0$;
	\item[(\romannumeral2)] the system
		\begin{equation}\label{eq:zero_init_cond}
		\left\{ \begin{array}{l}
		x_{e}(k+1) = \underbrace{\left[\Phi - \tilde E \left( H_\tau^f \right)^- C \Phi^\tau \right]}_{K_d} x_{e}(k) \\
		r_{e}(k) = \left[ I - H_\tau^f \left( H_\tau^f \right)^- \right] C  \Phi^\tau x_{e}(k) 
		\end{array} \right.
		\end{equation}
		is unobservable;
	\item[(\romannumeral3)] the inputs $\{f_{e}(i)\}$ take the form
		\begin{equation}\label{eq:zero_f}
		f_{e}(i) = - \left( H_\tau^f \right)^- C \Phi^\tau K_d^i x_{e}(0).
		\end{equation}
\end{enumerate}
\end{lem}

In Lemma \ref{lem:zero_dyn}, $r_{e}(0) = \cdots = r_{e}(\tau-1) = 0$ is ensured because of the condition (\romannumeral1) and the zero Markov matrices $H_0^f, H_1^f, \cdots, H_{\tau-1}^f$ according to Assumption \ref{ass:fault_rank}, while $r_{e}(\tau) = \cdots = r_{e}(L) = 0$ is ensured by the conditions (\romannumeral2) and (\romannumeral3).
Lemma \ref{lem:zero_dyn} can be proved by slightly modifying Lemmas A.1 and A.2 in \cite{Kirt2011}.

From Lemma \ref{lem:zero_dyn} we can see that perfect reconstruction of system inputs $\{f_e(i)\}$ from system outputs $\{r_e(i)\}$ is impossible if the unobservable input signal (\ref{eq:zero_f}) is non-zero. Hence, next, we will investigate the link between the unobservable input signal (\ref{eq:zero_f}) and the system property of $(\Phi, \tilde E, C, G )$.

By setting $i=0$, (\ref{eq:zero_f}) becomes 
\begin{equation}\label{eq:fe0}
f_{e}(0) = - \left( H_\tau^f \right)^- C \Phi^\tau x_{e}(0).
\end{equation}
Then, according to the condition (\romannumeral1) and the unobservability of the system (\ref{eq:zero_init_cond}), there must exist a scalar $\lambda$ and a non-zero $x_e(0)$ such that \cite{Zhoubook1996}
\begin{equation}\label{eq:inv_zero}
\begin{aligned}
\left[
\begin{smallmatrix}
K_d - \lambda I \\
\mathcal{O}_\tau \\
\left[ I - H_\tau^f \left( H_\tau^f \right)^- \right] C  \Phi^\tau
\end{smallmatrix}
\right] x_e(0) &= \left[ \begin{smallmatrix}
\Phi-\lambda I  & \tilde E \\
\mathcal{O}_\tau & 0 \\
C  \Phi^\tau & H_\tau^f
\end{smallmatrix} \right] \!
\left[ \begin{smallmatrix}
x_{e}(0) \\
f_{e}(0)
\end{smallmatrix} \right]  \\
&= \left[ \begin{smallmatrix}
\Phi-\lambda I  & \tilde E \\
\mathcal{O}_{\tau+1} & \mathbf{H}_\tau^f
\end{smallmatrix} \right] \!
\left[ \begin{smallmatrix}
x_{e}(0) \\
f_{e}(0)
\end{smallmatrix} \right] = 0,
\end{aligned}
\end{equation} 
where $\mathbf{H}_\tau^f$ defined in (\ref{eq:Htauf}) equals to $\left[\begin{smallmatrix}
0 \\ H_\tau^f
\end{smallmatrix}\right]$ because $H_0^f, H_1^f, \cdots, H_{\tau-1}^f$ are zero matrices according to Assumption \ref{ass:fault_rank}.
With (\ref{eq:fe0}) and $(K_d - \lambda I) x_e(0) = 0$ in (\ref{eq:inv_zero}), we can rewrite $f_e(i)$ in (\ref{eq:zero_f}) as
\begin{equation}\label{eq:fei}
f_e(i) = \lambda^i f_e(0). 
\end{equation}
The above analysis indicates that the unobservable inputs $\{f_e(i) = \lambda^i f_e(0)\}$ are determined by the invariant zero $\lambda$ of 
$( \Phi, \tilde E, \mathcal{O}_{\tau+1}, \mathbf{H}_{\tau}^f )$, as shown in the following lemma:

\begin{lem}\label{lem:null_space_zero}
Considering the non-zero initial state $x_e(0)$ in Lemma \ref{lem:zero_dyn}, there are two types of the invariant zeros $\lambda$ of the fault subsystem
$( \Phi, \tilde E, \mathcal{O}_{\tau+1}, \mathbf{H}_{\tau}^f )$ in (\ref{eq:inv_zero}): 1) 
$\lambda$ is an unobservable mode, then (\ref{eq:inv_zero}) implies $f_e(0)=0$, thus the input signal $\{f_e(i) = \lambda^i f_e(0)\}$ is constantly zero; 2) $\lambda$ is a transmission zero, then $f_e(0) \ne 0$, thus the  unobservable input signal $\{f_e(i) = \lambda^i f_e(0)\}$ is non-zero. 
\end{lem}

Lemma \ref{lem:null_space_zero} directly extends Lemmas 1 and 2 in \cite{Wan2014} which considers only the case $\tau = 0$.

\section{Proof of Theorem \ref{thm:unbias}}\label{app:thm_unbias}

A solution ${{\mathbf{\hat {{f}}}}_{k-\tau,L-\tau}^x}$ to the problem (\ref{eq:LS_prob}) satisfies
\begin{equation}\label{eq:LS_sol_cond}
\Psi_{L,\tau}^\mathrm{T} \Sigma_{e,L}^{-1} \Psi_{L,\tau} {{\mathbf{\hat {{f}}}}_{k-\tau,L-\tau}^x} =
\Psi_{L,\tau}^\mathrm{T} \Sigma_{e,L}^{-1} {{\bf{r}}_{k,L}}.
\end{equation}
Let $\Delta {{\mathbf{{{f}}}}_{k-\tau,L-\tau}^x} = {{\mathbf{\hat {{f}}}}_{k-\tau,L-\tau}^x} - {{\mathbf{{{f}}}}_{k-\tau,L-\tau}^x}$ denote the estimation error.
By substituting (\ref{eq:res_dyn_batch}) into (\ref{eq:LS_sol_cond}), we have
\begin{equation*}\label{eq:esterr}
\Psi_{L,\tau}^\mathrm{T} \Sigma_{e,L}^{-1} \Psi_{L,\tau} {\Delta {\mathbf{{{f}}}}_{k-\tau,L-\tau}^x} =
\Psi_L^\mathrm{T} \Sigma_{e,L}^{-1} {{\bf{e}}_{k,L}},
\end{equation*}
which implies
$\Psi_{L,\tau}^\mathrm{T} \Sigma_{e,L}^{-1} \Psi_{L,\tau} \mathrm{E}
\left( {\Delta {\mathbf{{{f}}}}_{k-\tau,L-\tau}^x} \right) = 0$
by taking expectations on both sides. Therefore, the unbiasedness condition of the estimate in (\ref{eq:f_hat_tao}) reduces to the analysis of the linear equation 
\begin{equation}\label{eq:fest_err_unbias}
	\Psi_{L,\tau}  \mathrm{E} \left( {\Delta {\mathbf{{{f}}}}_{k-\tau,L-\tau}^x} \right)= 0
\end{equation}
since 
$\mathcal{N}\left( \Psi_{L,\tau}^\mathrm{T} \Sigma_{e,L}^{-1} \Psi_{L,\tau} \right) = \mathcal{N}\left( \Psi_{L,\tau} \right)$.

The rest of the proof follows the intuitive arguments below.
According to Lemma \ref{lem:zero_dyn}, (\ref{eq:fei}), and the definition of $\mathbf{f}_{k-\tau,L-\tau}^x$ in (\ref{eq:res_dyn_batch}), there are three scenarios:
\begin{enumerate}
	\item[1)] When $( \Phi, \tilde E, \mathcal{O}_{\tau+1}, \mathbf{H}_{\tau}^f )$ has no invariant zeros, the non-zero initial state $x_e(0)$ in Lemma \ref{lem:zero_dyn} does not exist according to (\ref{eq:inv_zero}), thus (\ref{eq:fest_err_unbias}) implies $\mathrm{E} \left( {\Delta {\mathbf{{{f}}}}_{k-\tau,L-\tau}^x} \right)= 0$, i.e., unbiased fault estimation.
	\item[2)] When $( \Phi, \tilde E, \mathcal{O}_{\tau+1}, \mathbf{H}_{\tau}^f )$ has invariant zeros, (\ref{eq:fest_err_unbias}) implies that for each invariant zero $\lambda$, the expected error of the $\tau$-delay fault estimate $\hat f(k-\tau)$ is 
	\begin{equation}\label{eq:expct_fest_err}
	\text{E} \left(\Delta f(k-\tau)\right) = \lambda^{L-\tau-1} \text{E} \left(\Delta f(k_0)\right)
	\end{equation}
	in the estimation horizon $[k_0, k]$ ($k_0 = k-L+1$).
		\begin{enumerate}
			\item[2.1)] If all the invariant zeros of $( \Phi, \tilde E, \mathcal{O}_{\tau+1}, \mathbf{H}_{\tau}^f )$ correspond to unobservable modes, it follows from the case 1) in Lemma \ref{lem:null_space_zero} that the expected estimation error (\ref{eq:expct_fest_err}) is zero because $\text{E} \left(\Delta f(k_0)\right) = 0$.
			\item[2.2)] If transmission zeros exist but are all stable, i.e., $|\lambda| < 1$, it follows from the case 2) in Lemma \ref{lem:null_space_zero} that $\text{E} \left(\Delta f(k_0)\right) \ne 0$ and the expected estimation error (\ref{eq:expct_fest_err}) asymptotically reduced to zero as $L$ goes to infinity.
		\end{enumerate}
\end{enumerate}
The scenarios 1) and 2.1) correspond to the case (\romannumeral1) of Theorem \ref{thm:unbias}, and the scenario 2.2) corresponds to the case (\romannumeral2) of Theorem \ref{thm:unbias}.

\section{Proof of Theorem \ref{thm:solution_equivalent}}\label{app:equivalence}
For the original system model (\ref{eq:sys}), the extended output equation in the time window $\left[ k_0, k \right]$ is
\begin{equation}\label{eq:ext_output_origin}
\mathbf{y}_{k,L} = \mathscr{O}_{L} x(k_0) + \mathscr{T}_L^u \mathbf{u}_{k,L}
+ \mathscr{T}_{L}^{f} \mathbf{f}_{k,L} + \mathscr{T}_{L}^{w} \mathbf{w}_{k,L} + \mathbf{v}_{k,L},
\end{equation}
where $\mathscr{O}_L$, $\mathscr{T}_L^u$, $\mathscr{T}_L^f$, and $\mathscr{T}_L^w$ are defined
in the same way as $\mathcal{O}_L$ and $\mathbf{T}_L^u$ in (\ref{eq:OL_TLu}).
According to (\ref{eq:ext_output_origin}), we can rewrite (\ref{eq:resL_compute}) and (\ref{eq:res_dyn_batch0}) as
\begin{equation}\label{eq:res_compt_sys}
\begin{aligned}
\mathbf{r}_{k,L} &= \left( I - \mathbf{T}_L^y \right) \left( \mathbf{y}_{k,L} - \mathscr{T}_L^u \mathbf{u}_{k,L} \right) \\
&= \left( I - \mathbf{T}_L^y \right) \left( \mathscr{O}_{L} x(k_0) + \mathscr{T}_{L}^{f} \mathbf{f}_{k,L} + \mathscr{T}_{L}^{w} \mathbf{w}_{k,L} + \mathbf{v}_{k,L} \right)\\
&= \left( I - \mathbf{T}_L^y \right) \underbrace{\left[ \begin{array}{cc}
                                        \mathscr{O}_{L} & \mathscr{T}_{L,\tau}^{f}
                                      \end{array} \right]}_{\breve \Psi_{L,\tau}} \mathbf{f}_{k-\tau,L-\tau}^x  + \mathbf{e}_{k,L}.
\end{aligned}
\end{equation}
by following the relation between the original system model (\ref{eq:sys}) and its predictor form (\ref{eq:predictor}).
Similarly to $\mathbf{T}_{L,\tau}^f$ in (\ref{eq:res_dyn_batch}), $\mathscr{T}_{L,\tau}^{f}$ in (\ref{eq:res_compt_sys}) consists of the first $L-\tau$ block-columns of $\mathscr{T}_{L}^{f}$.

Define $\mathbf{\breve r}_{k,L} = \mathbf{y}_{k,L} - \mathscr{T}_L^u \mathbf{u}_{k,L}$ and
$${\breve \Sigma}_{L} = \rm{cov} \left( \mathscr{T}_{L}^{w} \mathbf{w}_{k,L} + \mathbf{v}_{k,L} \right).$$
Comparing (\ref{eq:res_dyn_batch}) with (\ref{eq:res_compt_sys}) leads to
\begin{equation}\label{eq:appC}
\begin{aligned}
&\mathbf{r}_{k,L} = \left( I - \mathbf{T}_L^y \right) \mathbf{\breve r}_{k,L}, \;\;
\Psi_{L,\tau} = \left( I - \mathbf{T}_L^y \right) {\breve \Psi}_{L,\tau}, \\
&\Sigma_{e,L} = \left( I - \mathbf{T}_L^y \right) {\breve \Sigma}_{L} \left( I - \mathbf{T}_L^y \right)^\mathrm{T}.
\end{aligned}
\end{equation}
Then by substituting (\ref{eq:appC}) into (\ref{eq:fkL_hat}), the estimate of ${{\mathbf{{{f}}}}_{k-\tau,L-\tau}^x}$ becomes
\begin{equation}\label{eq:fkL_hat_1}
{{\mathbf{\hat {{f}}}}_{k-\tau,L-\tau}^x} = \left( {\breve \Psi}_{L,\tau}^\mathrm{T} {\breve \Sigma}_{L}^{-1} {\breve \Psi}_{L,\tau} \right)^{(1)}
{\breve \Psi}_{L,\tau}^\mathrm{T} {\breve \Sigma}_{L}^{-1} {{\bf{\breve r}}_{k,L}},
\end{equation}
which is actually the LS estimate proposed in \cite{Wan2014} based on the original system model (\ref{eq:sys}).

\section{Proof of Theorem \ref{thm:unbias_markov}}\label{app:unbias_markov}

Split $\mathbf{T}_{L,\tau}^f$ into two blocks as
$\left[ \begin{array}{cc}
                                       \mathbf{\breve T}_{L,\tau}^{f} & \mathbf{\tilde T}_{L,\tau}^{f}
                                     \end{array} \right]$,
with $\mathbf{\breve T}_{L,\tau}^{f}$ consisting of the first $L-\tau-1$ block-columns of
$\mathbf{T}_{L,\tau}^f$, and $\mathbf{\tilde T}_{L,\tau}^{f}$ consisting of the last block-column of
$\mathbf{T}_{L,\tau}^f$. With these notations, unbiased fault estimation can be proved by showing that 
$\mathbf{\tilde T}_{L,\tau}^f \text{E}(\Delta f(k-\tau)) = 0$ because $\mathbf{\tilde T}_{L,\tau}^f$ has full column rank according to Assumption 
\ref{ass:fault_rank}.

According to (\ref{eq:range_HLm}), the following two expressions are equivalent:
\begin{align}
\varepsilon \in & \mathcal{R} \left( \left[ \begin{array}{cc}
\mathcal{O}_{L} & \mathbf{\breve T}_{L,\tau}^{f}
\end{array} \right] \right) \bigcap
\mathcal{R} \left( \mathbf{\tilde T}_{L,\tau}^{f} \right),  \label{eq:D1} \\
\varepsilon \in & \mathcal{R} \left( \left[ \begin{array}{cc}
\mathbf{H}_{L,m}^o & \mathbf{\breve T}_{L,\tau}^{f}
\end{array} \right] \right) \bigcap
\mathcal{R} \left( \mathbf{\tilde T}_{L,\tau}^{f} \right). \label{eq:D2}
\end{align}

\vspace{-0.5cm}

Since the two sufficient conditions for (asymptotic) unbiasedness in Theorem \ref{thm:unbias} imply $\varepsilon = 0$ and $\varepsilon \rightarrow 0$ ($L \rightarrow \infty$) for (\ref{eq:D1}), it then follows from the equivalence between (\ref{eq:D1}) and (\ref{eq:D2}) that 
the sufficient conditions in Theorem \ref{thm:unbias} also imply $\varepsilon = 0$ and $\varepsilon \rightarrow 0$ ($L \rightarrow \infty$) for (\ref{eq:D2}), or equivalently, $\mathcal{R} \left( \mathbf{\tilde T}_{L,\tau}^{f} \right) = \{0\}$ and $\mathcal{R} \left( \mathbf{\tilde T}_{L,\tau}^{f} \right) \rightarrow \{0\}$ ($L \rightarrow \infty$). Therefore we can conclude that the sufficient conditions in Theorem \ref{thm:unbias} imply  (asymptotically) unbiased fault estimation for (\ref{eq:D2}). Similarly, we can prove the necessary condition for the (asymptotically) unbiased fault estimation.

\section{{Computation of $\mathbb{\bar E} \left( {\mathcal{T}}_{s} \left( \mathcal{G} \right) {\mathcal{T}}_{s}^\mathrm{T} \left( \mathcal{G} \right) \right) $} } \label{app:computations}

By dividing $\mathbf{\bar M}_{\Upsilon}$ in (\ref{eq:Mbarf}) into $L$ row blocks as
\begin{equation}\label{eq:barMPsi_i}
\mathbf{\bar M}_{\Upsilon} = \left[ \begin{array}{cccc}
                                    \mathbf{M}_{\Upsilon,1}^\mathrm{T} & \mathbf{M}_{\Upsilon,2}^\mathrm{T} & \cdots & \mathbf{M}_{\Upsilon,L}^\mathrm{T}
                                   \end{array}
 \right]^\mathrm{T},
\end{equation}
with $\mathbf{M}_{\Upsilon,i} \in \mathbb{R}^{n_f \times \left( m\cdot n_u + \left( L-\tau \right)n_f \right)}$, we define $\mathbf{P}_{\Upsilon}$ as
\begin{equation}\label{eq:P_psi}
\mathbf{P}_{\Upsilon} = \left[ \begin{smallmatrix}
                             \mathrm{tr}\left( \mathbf{M}_{\Upsilon,1} \mathbf{M}_{\Upsilon,1}^\mathrm{T} \right) &
                             \mathrm{tr}\left( \mathbf{M}_{\Upsilon,1} \mathbf{M}_{\Upsilon,2}^\mathrm{T} \right)& \cdots & \mathrm{tr}\left( \mathbf{M}_{\Upsilon,1} \mathbf{M}_{\Upsilon,L}^\mathrm{T} \right) \\
                             \mathrm{tr}\left( \mathbf{M}_{\Upsilon,2} \mathbf{M}_{\Upsilon,1}^\mathrm{T} \right) &
                             \mathrm{tr}\left( \mathbf{M}_{\Upsilon,2} \mathbf{M}_{\Upsilon,2}^\mathrm{T} \right) & \cdots &
                             \mathrm{tr}\left( \mathbf{M}_{\Upsilon,2} \mathbf{M}_{\Upsilon,L}^\mathrm{T} \right) \\
                             \vdots & \vdots & \ddots & \vdots \\
                             \mathrm{tr}\left( \mathbf{M}_{\Upsilon,L} \mathbf{M}_{\Upsilon,1}^\mathrm{T} \right) &
                             \mathrm{tr}\left( \mathbf{M}_{\Upsilon,L} \mathbf{M}_{\Upsilon,2}^\mathrm{T} \right) &
                             \cdots & \mathrm{tr}\left( \mathbf{M}_{\Upsilon,L} \mathbf{M}_{\Upsilon,L}^\mathrm{T} \right)
                           \end{smallmatrix}
 \right].
\end{equation}
$\mathbf{P}_z$ is defined similarly to (\ref{eq:P_psi}), by dividing $\mathbf{\bar M}_L^z$ in (\ref{eq:hatrkL}) into $L$ row blocks as in (\ref{eq:barMPsi_i}). Then,
\begin{gather}
\mathbb{\bar E} \left( {\mathcal{T}}_{f} \left( \mathcal{G} \right) {\mathcal{T}}_{f}^\mathrm{T} \left( \mathcal{G} \right) \right)
= \left[ \begin{array}{cc}
         \mathcal{G} & \mathcal{I}_{n_f}
       \end{array} \right]
\left[ \begin{array}{cc}
         \Pi_f & -\hat {\Upsilon}_{L,\tau} \\
         -\hat {\Upsilon}_{L,\tau}^\mathrm{T} & I_{n_f}
       \end{array}
 \right]
\left[ \begin{array}{c}
         \mathcal{G}^\mathrm{T} \\
         \mathcal{I}_{n_f}^\mathrm{T}
       \end{array}
 \right], \\
\mathbb{\bar E} \left( {\mathcal{T}}_{z} \left( \mathcal{G} \right) {\mathcal{T}}_{z}^\mathrm{T} \left( \mathcal{G} \right) \right)
= \mathcal{G} \Pi_z \mathcal{G}^\mathrm{T}
\end{gather}
with
\begin{equation}\label{eq:pif}
\begin{aligned}
\Pi_f =&\, \hat {\Upsilon}_{L,\tau} \hat {\Upsilon}_{L,\tau}^\mathrm{T} +
         \mathbb{\bar E}\left( \mathbf{\bar E}_{\mathrm{id}} \mathbf{\bar M}_{\Upsilon} \mathbf{\bar M}_{\Upsilon}^\mathrm{T} \mathbf{\bar E}_{\mathrm{id}}^\mathrm{T} \right) \\
      =&\, \hat {\Upsilon}_{L,\tau} \hat {\Upsilon}_{L,\tau}^\mathrm{T} + \mathbf{P}_{\Upsilon} \otimes \Sigma_e,
\end{aligned}
\end{equation}
\begin{equation}\label{eq:piz}
\Pi_z = \mathbb{\bar E}\left( \mathbf{\bar E}_{\mathrm{id}} \mathbf{\bar M}_{L}^z (\mathbf{\bar M}_{L}^z)^{\mathrm{T}} \mathbf{\bar E}_{\mathrm{id}}^\mathrm{T} \right) = \mathbf{P}_{z} \otimes \Sigma_e.
\end{equation}

\end{document}